\documentclass[prd,preprint,tightenlines,floatfix,
preprintnumbers,nofootinbib,eqsecnum,superscriptaddress]{revtex4}

\usepackage[T1]{fontenc}		

\usepackage{amsmath,amsfonts,amssymb,amstext,mathrsfs}
\usepackage{mathpazo}

\usepackage[dvips]{graphicx}
\usepackage{epsf,float}
\usepackage{revsymb}

\usepackage{dcolumn}
\usepackage{braket}
\usepackage{color,xcolor}
\usepackage{graphicx}
\usepackage{subfigure}
\usepackage{multirow}
\usepackage{tabularx}
\usepackage{pstricks}
\usepackage[section]{placeins}
\usepackage{booktabs}
\usepackage{array}

\usepackage{hyperref}

\newcommand{\Pom}{\mathbb{P}}
\newcommand{\Ode}{\mathbb{O}}
\newcommand{\Reg}{\mathbb{R}}

\renewcommand\slash[1]{\not \! #1}

\begin{document}

\title{Central exclusive diffractive production of \boldmath{$p \bar{p}$} pairs\\
in proton-proton collisions at high energies}

\author{Piotr Lebiedowicz}
 \email{Piotr.Lebiedowicz@ifj.edu.pl}
\affiliation{Institute of Nuclear Physics Polish Academy of Sciences, Radzikowskiego 152, PL-31-342 Krak\'ow, Poland}

\author{Otto Nachtmann}
 \email{O.Nachtmann@thphys.uni-heidelberg.de}
\affiliation{Institut f\"ur Theoretische Physik, Universit\"at Heidelberg,
Philosophenweg 16, D-69120 Heidelberg, Germany}

\author{Antoni Szczurek
\footnote{Also at \textit{Faculty of Mathematics and Natural Sciences, University of Rzesz\'ow, Pigonia 1, PL-35-310 Rzesz\'ow, Poland}.}}
\email{Antoni.Szczurek@ifj.edu.pl}
\affiliation{Institute of Nuclear Physics Polish Academy of Sciences, Radzikowskiego 152, PL-31-342 Krak\'ow, Poland}

\begin{abstract}
We consider the central exclusive production of the $p\bar{p}$ in the continuum
and via resonances in proton-proton collisions at high energies.
We discuss the diffractive mechanism calculated within 
the tensor-pomeron approach including pomeron, odderon, and reggeon exchanges.
The theoretical results are discussed in the context of existing WA102 and ISR experimental data
and predictions for planned or current experiments 
at the RHIC and the LHC are presented.
The distribution in ${\rm y}_{diff}$, 
the rapidity distance between proton and antiproton,
is particularly interesting.
We find a dip at ${\rm y}_{diff} = 0$ for the $p \bar{p}$ production,
in contrast to the $\pi^{+}\pi^{-}$ and $K^{+}K^{-}$ production.
We predict also the $p \bar{p}$ invariant mass distribution
to be less steep than for the pairs of pseudoscalar mesons.
We argue that these specific differences for the $p \bar{p}$ production with respect
to the pseudoscalar meson pair production 
can be attributed to the proper treatment of the spin of produced particles.
We discuss asymmetries that are due to the interference
of $C=+1$ and $C=-1$ amplitudes of $p \bar{p}$ production.
We have also calculated the cross section for 
the $pp \to pp \Lambda \overline{\Lambda}$ reaction.
Here, the cross section is smaller but
the characteristic feature for $d\sigma/d{\rm y}_{diff}$ is predicted
to be similar to $p \bar{p}$ production.
The presence of resonances in the $p \bar p$ channel
may destroy the dip at ${\rm y}_{diff} = 0$.
This opens the possibility to study diffractively produced resonances.
We discuss the observables suited for this purpose.
\end{abstract}

\maketitle

\section{Introduction}
\label{sec:intro}
Diffractive exclusive production of resonances and of dihadron continua
are processes with relatively large cross sections, 
typically of the order of a few $\mu$b or even larger.
It~is~expected that central exclusive production,
mediated by double pomeron exchange,
is an ideal reaction for the investigation of gluonic bound states (glueballs)
of which the existence has not yet been confirmed unambiguously.
Observation of glueballs would be a long-awaited confirmation 
of a crucial prediction of the QCD theory.
Such processes were studied extensively at CERN starting from 
the Intersecting Storage Rings (ISR) experiments
\cite{Waldi:1983sc,Akesson:1985rn,Breakstone:1986xd,
Breakstone:1989ty,Breakstone:1990at}
(for a review, see Ref.~\cite{Albrow:2010yb})
and later at the Omega spectrometer at SPS 
in the fixed-target WA102 experiment; 
see e.g.~\cite{Barberis:1996iq,Barberis:1997ve,Barberis:1998ax,
Barberis:1998sr,Barberis:1999cq,Barberis:2000em,Kirk:2000ws}.
The measurement of two charged pions in $p \bar{p}$ collisions
was performed by the CDF Collaboration at the Tevatron \cite{Aaltonen:2015uva}.
In this experiment the outgoing $p$ and $\bar{p}$ were not detected
and only two large rapidity gaps, one on each side of the central hadronic system, were required.
Thus, the data include also diffractive dissociation of (anti)protons
into undetected hadrons.
Exclusive reactions are of particular interest
since they can be studied in experiments at the LHC
by the ALICE, ATLAS, CMS \cite{Khachatryan:2017xsi}, and LHCb collaborations.
At the LHC, in the reactions of interest here,
protons are scattered in the forward/backward
directions in which relevant detectors are not always present.
Recently, there have been several efforts to install and use forward proton detectors.  
The CMS Collaboration combines efforts with the TOTEM Collaboration 
while the ATLAS Collaboration may use the ALFA subdetectors; see e.g. \cite{Staszewski:2011bg}.
Also the STAR experiment at the Relativistic Heavy Ion Collider (RHIC) 
is equipped with such detectors 
that allow the measurement of forward protons.
In this way, the non-exclusive background due to proton breakup
could be rejected via the momentum balance constraint \cite{Adamczyk:2014ofa,Sikora:2016evz}.

On the theoretical side, the exclusive diffractive dihadron continuum production 
can be understood as being mainly due to the exchange of two pomerons
between the external protons and the centrally produced hadronic system.
First calculations in this respect were concerned 
with the $p p \to p p \pi^+ \pi^-$ reaction 
\cite{Pumplin:1976dm,Lebiedowicz:2009pj,Lebiedowicz:2011nb}.
The Born amplitude was written in terms of pomeron/reggeon exchanges 
with parameters fixed from phenomenological analyses of
$NN$ and $\pi N$ total and elastic scattering.
The four-body amplitude was parametrized using 
the four-momentum transfers squared
$t_1$, $t_2$, and $s_{ij}$, the energies squared in the two-body subsystems.
The energy dependence is known from two-body scatterings such as 
$NN$, $\pi N$, etc.
Such calculations make sense for the continuum production
of pseudoscalar meson pairs.
These model studies were extended also to
the $pp \to nn \pi^+ \pi^+$ \cite{Lebiedowicz:2010yb} and
$pp \to pp K^+ K^-$ \cite{Lebiedowicz:2011tp} reactions
and even for the exclusive $\pi^{+}\pi^{-}\pi^{+}\pi^{-}$ 
continuum production \cite{Kycia:2017iij}.
In reality the Born approximation is usually not sufficient and 
absorption corrections have to be taken into account;
see e.g. \cite{Harland-Lang:2013dia,Lebiedowicz:2015eka}.
The phenomenological concepts underlying these calculations
require further tests and clear
phenomenological evidence to be commonly accepted.

In this paper we are concerned with reactions in which the exchange
of the soft pomeron plays the most important role.
This -- still somewhat enigmatic -- soft pomeron is a flavorless object. 
It is often loosely stated that it possesses quantum numbers of the vacuum.
This is true for the internal quantum numbers of the pomeron.
However, the spin structure of the soft pomeron certainly is
not that of the vacuum, i.e. spin 0.
We believe that the soft pomeron is best described as an effective
rank-2 symmetric-tensor exchange as introduced in \cite{Ewerz:2013kda}.
In \cite{Ewerz:2016onn} three hypotheses for the soft-pomeron spin structure,
effective scalar, vector, and tensor exchange, were discussed and
compared to the experimental data on the helicity structure of proton-proton elastic scattering at
$\sqrt{s} = 200$~GeV and small $|t|$ from the STAR experiment \cite{Adamczyk:2012kn}.
Only the tensor option was shown to be viable,
the vector and scalar options for the soft pomeron could be excluded.
In \cite{Ewerz:2016onn} also some remarks on the history 
of the views of the pomeron spin structure were presented.
For the convenience of the reader we repeat here some of the main points concerning
the tensor pomeron in its connection to QCD.
In \cite{Nachtmann:1991ua}, one of us made a general investigation of high-energy soft diffractive
processes in QCD using functional methods.
It was shown there that the resulting soft pomeron could be described as coherent exchange
of spin $2+4+6+\ldots$. 
This is exactly the structure of the tensor pomeron of \cite{Ewerz:2013kda};
see Appendix~B there.
In this way the tensor pomeron of \cite{Ewerz:2013kda} has good backing in 
nonperturbative QCD. Also investigations in the framework of the AdS/CFT correspondence
prefer a tensor nature for the soft-pomeron exchange \cite{Domokos:2009hm,Iatrakis:2016rvj}.

First applications of the tensor-pomeron model of \cite{Ewerz:2013kda}
to the central exclusive production (CEP) of several scalar and pseudoscalar mesons 
in the reaction $p p \to p p M$ were studied in \cite{Lebiedowicz:2013ika}
for the relatively low WA102 energy, where also the secondary reggeon exchanges
play a very important role.
The resonant $\rho^0$ ($J^{PC} = 1^{--}$) and non-resonant (Drell-S\"oding)
$\pi^{+}\pi^{-}$ photoproduction contributions to CEP were studied in \cite{Lebiedowicz:2014bea}.
In \cite{Bolz:2014mya}, an extensive study of the reaction 
$\gamma p \to \pi^+ \pi^- p$ was presented.
In \cite{Lebiedowicz:2016ioh}, the model was applied
to the reaction $p p \to p p \pi^+ \pi^-$
including the dipion continuum, the dominant scalar
$f_{0}(500)$, $f_{0}(980)$~($J^{PC} = 0^{++}$),
and tensor $f_{2}(1270)$~($J^{PC} = 2^{++}$) resonances 
decaying into the $\pi^+ \pi^-$ pairs.
In \cite{Lebiedowicz:2016zka}, the model was applied
to the $\pi^+ \pi^-\pi^+ \pi^-$ production 
via the intermediate $\sigma\sigma$ and $\rho^0\rho^0$ states.
Also the $\rho^{0}$ meson production associated 
with a very forward/backward $\pi N$ system,
that is, the $pp \to pp \rho^{0} \pi^{0}$ and $pp \to pn \rho^{0} \pi^{+}$ processes
were discussed in \cite{Lebiedowicz:2016ryp}.
It was shown in 
\cite{Lebiedowicz:2013ika,Lebiedowicz:2016ioh,Lebiedowicz:2014bea,Bolz:2014mya,
Lebiedowicz:2016ryp,Lebiedowicz:2016zka}
that the tensor-pomeron model does 
quite well in reproducing the data where available.

Closely related to the reaction $pp \to pp p \bar{p}$ studied by us here
are the reactions of central $p \bar{p}$ production in
ultraperipheral nucleus-nucleus and nucleon-nucleus collisions,
$AA \to AA p \bar{p}$ and $pA \to pA p \bar{p}$.
For the first process, see \cite{Klusek-Gawenda:2017lgt},
in which the parameters of the model including the proton exchange, 
the $f_{2}(1270)$ and $f_{2}(1950)$ $s$-channel exchanges, 
and the handbag mechanism,
were fitted to Belle data \cite{Kuo:2005nr} 
for the $\gamma \gamma \to p \bar{p}$ reaction.
The model was applied then to estimate the cross section for 
the ultraperipheral, ultrarelativistic, heavy-ion collisions at the LHC.

In the following, we extend the application of the tensor-pomeron model to central exclusive
production of spin-$1/2$ hadron pairs ($p \bar{p}$ or 
$\Lambda \overline{\Lambda}$) in $pp$ collisions. 
The centrally produced baryon-antibaryon pairs were studied experimentally 
in Refs.~\cite{Akesson:1985rn,Breakstone:1989ty, Barberis:1998sr}.
So far, the $p p \to p p p \bar{p}$ reaction 
at LHC energies has not been considered
from the theory point of view.
We will show first predictions for this reaction
in the tensor-pomeron approach
and compare them with results for central production
of dihadrons with spin 0, $\pi^{+} \pi^{-}$ and $K^{+} K^{-}$.
We shall discuss whether the predictions of the tensor-pomeron model 
can be verified by planned measurements at the RHIC and at the LHC. 
The observables suited for this purpose shall be presented.

Our paper is organised as follows.
In Sec.~\ref{sec:continuum_ppbar} we discuss continuum $p \bar{p}$ production.
Section~\ref{sec:section_f0_ppbar} deals with $p \bar{p}$ production
via scalar resonances. First results are presented in
Sec.~\ref{sec:results}, and Sec.~\ref{sec:conclusions}
presents our conclusions. We include in our calculations
the exchanges of the soft pomeron, of reggeons,
and also of the soft odderon for some distributions.
The odderon was introduced a long time ago \cite{Lukaszuk:1973nt,Joynson:1975az}
(for a review, see, e.g. \cite{Ewerz:2003xi}) and has recently become very interesting again 
\cite{Antchev:2017yns,Martynov:2017zjz,Khoze:2017swe,Khoze:2018bus}.

We want to emphasise that the purpose of our paper
is not to compare predictions of our tensor-pomeron approach
with alternatives for the soft-pomeron structure.
This has been done extensively in \cite{Lebiedowicz:2013ika,Ewerz:2016onn}.
Also, since we are interested in the soft-scattering regime
we cannot use or compare with the perturbative pomeron, 
initiated in \cite{Low:1975sv,Nussinov:1975mw,Kuraev:1977fs,Balitsky:1978ic}.
The purpose of our work is to give experimentalists
a solid idea of what to expect theoretically in central exclusive $p \bar{p}$ production.
What are the magnitudes of cross sections? Where is continuum $p \bar{p}$?
Where is resonance production prominent? What is the role of secondary reggeon exchanges
and, if it exists, of odderon exchange?
What are the differences between $p \bar{p}$ and two pseudoscalars central production?
We would hope that our calculations could serve as basis 
for the construction of an event generator for this and related processes.
\footnote{The $\mathtt{GenEx}$ Monte Carlo generator \cite{Kycia:2014hea} 
could be used and expanded in this context.}
A long-term goal would be to derive the coupling constants of our effective
theory from nonperturbative QCD, but this is beyond the scope of the present paper.

\section{$p \bar{p}$ continuum production}
\label{sec:continuum_ppbar}

We study central exclusive production of $p \bar{p}$ 
in proton-proton collisions at high energies
\begin{eqnarray}
p(p_{a},\lambda_{a}) + p(p_{b},\lambda_{b}) \to
p(p_{1},\lambda_{1}) + \bar{p}(p_{3},\lambda_{3}) + p(p_{4},\lambda_{4}) + p(p_{2},\lambda_{2}) \,,
\label{2to4_reaction}
\end{eqnarray}
where $p_{i}$ and 
$\lambda_{i} \in \lbrace +1/2, -1/2 \rbrace$, indicated in brackets,
denote the 4-momenta and helicities of the nucleons, respectively.
The ${\cal T}$-matrix element for the reaction (\ref{2to4_reaction})
will be denoted as follows
\begin{equation}
\begin{split}
{\cal M}_{\lambda_{a}\lambda_{b} \to \lambda_{1}\lambda_{2}\lambda_{3}\lambda_{4}} =
\langle 
p(p_{1}, \lambda_{1}), p(p_{4}, \lambda_{4}),
\bar{p}(p_{3}, \lambda_{3}) p(p_{2}, \lambda_{2}) | {\cal T} &| 
p(p_{a}, \lambda_{a}), p(p_{b}, \lambda_{b})\rangle\,.
\end{split}
\label{Tamplitude}
\end{equation}
Note that the order of the particles in the bra and ket states matters
since we are dealing with fermions.

In general the full amplitude for the $p\bar{p}$ production 
is a sum of the continuum amplitude 
and the amplitudes with the $s$-channel resonances:
\begin{equation}
\begin{split}
{\cal M}_{\lambda_{a}\lambda_{b} \to \lambda_{1}\lambda_{2}\lambda_{3}\lambda_{4}} =
{\cal M}^{p\bar{p}{\rm-continuum}}_{\lambda_{a}\lambda_{b} \to \lambda_{1}\lambda_{2}\lambda_{3}\lambda_{4}} + 
{\cal M}^{p\bar{p}{\rm-resonances}}_{\lambda_{a}\lambda_{b} \to \lambda_{1}\lambda_{2}\lambda_{3}\lambda_{4}}\,.
\end{split}
\label{amplitude_pomTpomT}
\end{equation}
At high energies the exchange objects to be considered are
the photon $\gamma$, the pomeron $\Pom$, the odderon $\Ode$, 
and the reggeons $\Reg$.
Their charge conjugation and $G$-parity quantum numbers
are listed in Table~I of \cite{Lebiedowicz:2016ioh}.
We treat the $C=+1$ pomeron and the reggeons $\Reg_{+} = f_{2 \Reg}, a_{2 \Reg}$ 
as effective tensor exchanges
while the $C=-1$ odderon and the reggeons 
$\Reg_{-} = \omega_{\Reg}, \rho_{\Reg}$ are treated as effective vector exchanges.

\begin{figure}
\includegraphics[width=6.cm]{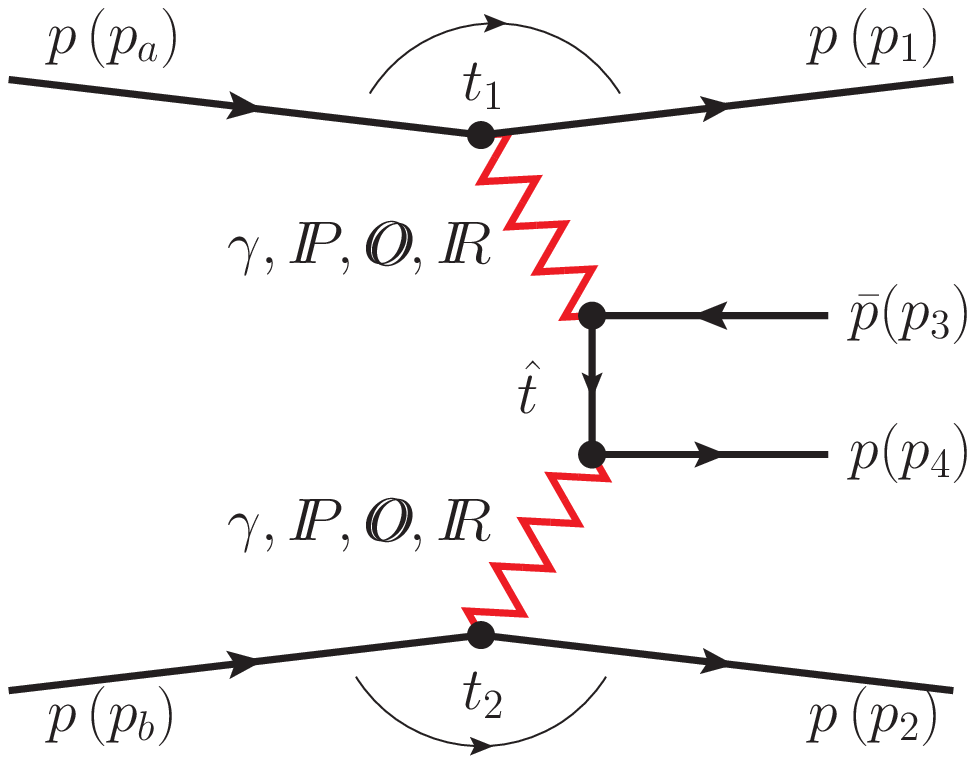}    
\includegraphics[width=6.cm]{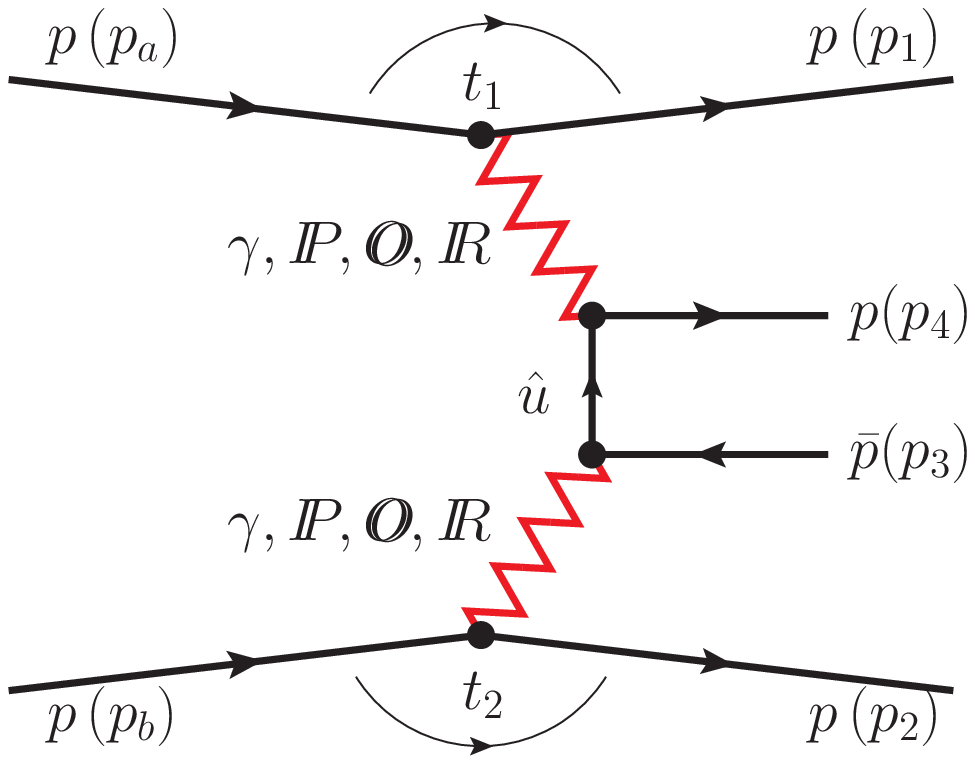}     
  \caption{\label{fig:born_diagrams}
  \small The Born diagrams for the double pomeron/reggeon
  and photon-mediated central exclusive continuum 
  $p\bar{p}$ production in proton-proton collisions.
}
\end{figure}
The $p \bar{p}$ continuum amplitude
is expressed as the sum of ${\rm \hat{t}}$ and ${\rm \hat{u}}$ diagrams 
shown in Fig.~\ref{fig:born_diagrams},
\begin{eqnarray}
{\cal M}^{p\bar{p}-\rm{continuum}}_{\lambda_{a} \lambda_{b} \to \lambda_{1} \lambda_{2} \lambda_{3} \lambda_{4}} =
{\cal M}^{({\rm \hat{t}})}_{\lambda_{a} \lambda_{b} \to \lambda_{1} \lambda_{2} \lambda_{3} \lambda_{4}}+
{\cal M}^{({\rm \hat{u}})}_{\lambda_{a} \lambda_{b} \to \lambda_{1} \lambda_{2} \lambda_{3} \lambda_{4}}
\,.
\label{pompom_amp}
\end{eqnarray}
The combinations $(C_{1}, C_{2})$ of exchanges that can contribute 
in (\ref{pompom_amp}) are
\begin{eqnarray}
&& (C_{1}, C_{2}) = (1, 1): \;(\Pom + \Reg_{+}, \Pom + \Reg_{+})\,;
\label{C_modes_ppbar_pp}\\
&& (C_{1}, C_{2}) = (-1, -1):\; (\Ode + \Reg_{-} + \gamma, \Ode + \Reg_{-} + \gamma)\,;
\label{C_modes_ppbar_mm}\\
&& (C_{1}, C_{2}) = (1, -1):\; (\Pom + \Reg_{+}, \Ode + \Reg_{-} + \gamma)\,;
\label{C_modes_ppbar_pm}\\
&& (C_{1}, C_{2}) = (-1, 1):\; (\Ode + \Reg_{-} + \gamma, \Pom + \Reg_{+})\,.
\label{C_modes_ppbar_mp}
\end{eqnarray}
Here $C_{1}$ and $C_{2}$ are the charge-conjugation
quantum numbers of the exchange objects.
The contributions involving the photon $\gamma$
in (\ref{C_modes_ppbar_mm}) to (\ref{C_modes_ppbar_mp}) are expected to be small
but may be important at very small four-momentum transfer squared.
The $(\gamma, \Pom + \Reg_{+})$ and $(\Pom + \Reg_{+}, \gamma)$ contributions
will be very important for the $p \bar{p}$ production in $pA$ collisions.
There one also has to take into account contact terms required by
gauge invariance.
This will be studied elsewhere.
The contributions involving the odderon $\Ode$ are expected to be small 
since its coupling to the proton is very small.
Thus, we shall concentrate on the diffractive production of $p \bar{p}$
through the $\Pom$, $\Reg_{+}$, and $\Reg_{-}$ exchanges
but also mention odderon effects where appropriate.

The kinematic variables for reaction (\ref{2to4_reaction}) are
\begin{eqnarray}
&&s = (p_{a} + p_{b})^{2} = (p_{1} + p_{2} + p_{3} + p_{4})^{2}\,,
\nonumber \\ 
&&s_{ij} = (p_{i} + p_{j})^{2}, \quad s_{34} = M_{34}^{2} = (p_{3} + p_{4})^{2}\,,\nonumber \\
&&t_1 = q_{1}^{2}, \quad q_1 = p_{a} - p_{1}\,,
\nonumber \\
&&t_2 = q_{2}^{2}, \quad q_2 = p_{b} - p_{2}\,, 
\nonumber \\
&&
\hat{p}_{t} = p_{a} - p_{1} - p_{3}\,,
\nonumber \\
&&
\hat{p}_{u} = p_{4} - p_{a} + p_{1}\,.
\label{2to4_kinematic}
\end{eqnarray}

Let us first take a look at the dominant ($\Pom$, $\Pom$) contribution.
The ${\rm \hat{t}}$- and ${\rm \hat{u}}$-channel amplitudes 
for the $\Pom \Pom$-exchange can be written as
\begin{equation}
\begin{split}
& {\cal M}^{({\rm \hat{t}})}_{\lambda_{a} \lambda_{b} 
\to \lambda_{1} \lambda_{2} \lambda_{3} \lambda_{4}} 
= (-i)
\bar{u}(p_{1}, \lambda_{1}) 
i\Gamma^{(\Pom pp)}_{\mu_{1} \nu_{1}}(p_{1},p_{a}) 
u(p_{a}, \lambda_{a})\;
i\Delta^{(\Pom)\, \mu_{1} \nu_{1}, \alpha_{1} \beta_{1}}(s_{13},t_{1})\\
& \qquad \qquad \qquad \quad \times 
\bar{u}(p_{4}, \lambda_{4}) \, 
i\Gamma^{(\Pom pp)}_{\alpha_{2} \beta_{2}}(p_{4},\hat{p}_{t}) \,
iS_{F}(\hat{p}_{t})\,
i\Gamma^{(\Pom pp)}_{\alpha_{1} \beta_{1}}(\hat{p}_{t},-p_{3})\,
v(p_{3}, \lambda_{3}) \\
& \qquad \qquad \qquad \quad \times 
i\Delta^{(\Pom)\, \alpha_{2} \beta_{2}, \mu_{2} \nu_{2}}(s_{24},t_{2})\,
\bar{u}(p_{2}, \lambda_{2}) 
i\Gamma^{(\Pom pp)}_{\mu_{2} \nu_{2}}(p_{2},p_{b}) 
u(p_{b}, \lambda_{b}) \,,
\end{split}
\label{amplitude_pompom_t}
\end{equation}
\begin{equation}
\begin{split}
& {\cal M}^{({\rm \hat{u}})}_{\lambda_{a} \lambda_{b} 
\to \lambda_{1} \lambda_{2} \lambda_{3} \lambda_{4}} 
= (-i)
\bar{u}(p_{1}, \lambda_{1}) 
i\Gamma^{(\Pom pp)}_{\mu_{1} \nu_{1}}(p_{1},p_{a}) 
u(p_{a}, \lambda_{a})\;
i\Delta^{(\Pom)\, \mu_{1} \nu_{1}, \alpha_{1} \beta_{1}}(s_{14},t_{1})\\
& \qquad \qquad \qquad \quad \times 
\bar{u}(p_{4}, \lambda_{4}) \, 
i\Gamma^{(\Pom pp)}_{\alpha_{1} \beta_{1}}(p_{4},\hat{p}_{u})\,
iS_{F}(\hat{p}_{u}) \, 
i\Gamma^{(\Pom pp)}_{\alpha_{2} \beta_{2}}(\hat{p}_{u},-p_{3})\,
v(p_{3}, \lambda_{3}) \\
& \qquad \qquad \qquad \quad \times 
i\Delta^{(\Pom)\, \alpha_{2} \beta_{2}, \mu_{2} \nu_{2}}(s_{23},t_{2})\,
\bar{u}(p_{2}, \lambda_{2}) 
i\Gamma^{(\Pom pp)}_{\mu_{2} \nu_{2}}(p_{2},p_{b}) 
u(p_{b}, \lambda_{b}) \,.
\end{split}
\label{amplitude_pompom_u}
\end{equation}
Here we use the standard propagator for the proton
$i S_{F}(\hat{p}) = i / (\hat{p}\!\!\!/ - m_{p})$.
The effective propagator of the tensor-pomeron exchange 
and the pomeron-proton vertex function are given in section~3 of \cite{Ewerz:2013kda}.
For the convenience of the reader we collect these and other quantities which we use
in our work in Appendix~\ref{sec:appendixA}.

For $\Pom \Pom$ fusion the centrally produced $p \bar{p}$ system
is in a state of $C = +1$.
This implies for the amplitude (\ref{Tamplitude}) the following:
\begin{equation}
\begin{split}
&\langle 
p(p_{1}, \lambda_{1}), p(p_{4}, \lambda_{4}),
\bar{p}(p_{3}, \lambda_{3}), p(p_{2}, \lambda_{2}) | {\cal T} | 
p(p_{a}, \lambda_{a}), p(p_{b}, \lambda_{b})\rangle^{(\Pom,\Pom)} \\
&=\langle 
p(p_{1}, \lambda_{1}), \bar{p}(p_{4}, \lambda_{4}),
p(p_{3}, \lambda_{3}), p(p_{2}, \lambda_{2}) | {\cal T} | 
p(p_{a}, \lambda_{a}), p(p_{b}, \lambda_{b})\rangle^{(\Pom,\Pom)} \\
&=-\langle 
p(p_{1}, \lambda_{1}), p(p_{3}, \lambda_{3}),
\bar{p}(p_{4}, \lambda_{4}), p(p_{2}, \lambda_{2}) | {\cal T} | 
p(p_{a}, \lambda_{a}), p(p_{b}, \lambda_{b})\rangle^{(\Pom,\Pom)} \,.
\end{split}
\label{Tamplitude_aux}
\end{equation}
Here, we work in the overall c.m. system and 
assume that the helicity states for the centrally produced $p$ and $\bar{p}$
are both taken of the same type, e.g., of type (a); 
see Appendix~A of \cite{Klusek-Gawenda:2017lgt}.
This antisymmetry relation (\ref{Tamplitude_aux}) can,
of course, be verified explicitly using the expressions for
${\cal M}^{({\rm \hat{t}})}$ and ${\cal M}^{({\rm \hat{u}})}$
from (\ref{amplitude_pompom_t}) and (\ref{amplitude_pompom_u}), respectively.

If we use another choice of $p$ and $\bar{p}$ helicity states in the c.m. system
we will get additional phase factors in (\ref{Tamplitude_aux})
and the corresponding relations for the other $(C_{1}, C_{2})$ exchanges.
But these phase factors drop out for distributions
where the polarisations of the centrally produced $p$ and $\bar{p}$
are not observed.
Thus, our above choice for the $p$ and $\bar{p}$ helicity states
is very convenient as it makes the $p \bar{p}$ charge-conjugation relations
for the amplitudes simple and explicit.

The antisymmetry relation (\ref{Tamplitude_aux}) holds for all
exchanges with $(C_{1}, C_{2})$ = $(1,1)$ and $(-1,-1)$;
see (\ref{C_modes_ppbar_pp}) and (\ref{C_modes_ppbar_mm}).
For the exchanges with $(C_{1}, C_{2})$ = $(1,-1)$ and $(-1,1)$ we have,
instead, symmetry under the exchange
$\left( p(p_{4}, \lambda_{4}), \bar{p}(p_{3}, \lambda_{3}) \right) \to 
\left( p(p_{3}, \lambda_{3}), \bar{p}(p_{4}, \lambda_{4}) \right)$;
see (\ref{C_modes_ppbar_pm}) and (\ref{C_modes_ppbar_mp}).

In the high-energy approximation,
we can write the $\Pom \Pom$-exchange amplitude as
\begin{equation}
\begin{split}
& {\cal M}^{(\Pom \Pom \to p \bar{p})}_{\lambda_{a} \lambda_{b} \to 
\lambda_{1} \lambda_{2} \lambda_{3} \lambda_{4}} 
\simeq
(3 \beta_{\Pom NN})^{2}\, 2 (p_1 + p_a)_{\mu_{1}} (p_1 + p_a)_{\nu_{1}}\, 
\delta_{\lambda_{1} \lambda_{a}} \,[F_{1}(t_{1})]^{2}\\
& \qquad\qquad\qquad\quad \times  
\bar{u}(p_{4}, \lambda_{4}) 
\Large[
\gamma^{\mu_{2}}(p_{4}+\hat{p}_{t})^{\nu_{2}} \,
\frac{1}{4 s_{13}} (- i s_{13} \alpha'_{\Pom})^{\alpha_{\Pom}(t_{1})-1}
\frac{[\hat{F}_{p}(\hat{p}_{t}^{2})]^{2}}{\slash{\hat{p}}_{t} - m_{p}}\\
& \qquad\qquad\qquad \quad    \times  
\gamma^{\mu_{1}}(\hat{p}_{t}-p_{3})^{\nu_{1}} \,
\frac{1}{4 s_{24}} (- i s_{24} \alpha'_{\Pom})^{\alpha_{\Pom}(t_{2})-1}\\
&\qquad\qquad\qquad \qquad   +
\gamma^{\mu_{1}}(p_{4}+\hat{p}_{u})^{\nu_{1}} 
\frac{1}{4 s_{14}} (- i s_{14} \alpha'_{\Pom})^{\alpha_{\Pom}(t_{1})-1}
\frac{[\hat{F}_{p}(\hat{p}_{u}^{2})]^{2}}{\slash{\hat{p}}_{u} - m_{p}} \\
& \qquad\qquad\qquad \quad   \times  
\gamma^{\mu_{2}}(\hat{p}_{u}-p_{3})^{\nu_{2}} \,
\frac{1}{4 s_{23}} (- i s_{23} \alpha'_{\Pom})^{\alpha_{\Pom}(t_{2})-1}
\Large]\,
v(p_{3}, \lambda_{3}) \\
& \qquad\qquad\qquad\quad \times  
(3 \beta_{\Pom NN})^{2}\,2 (p_2 + p_b)_{\mu_{2}} (p_2 + p_b)_{\nu_{2}}\, 
\delta_{\lambda_{2} \lambda_{b}}\, [F_{1}(t_{2})]^{2} \,.
\end{split}
\label{amplitude_pompom_approx}
\end{equation}
In (\ref{amplitude_pompom_approx}), we have introduced a form factor
$\hat{F}_{p}(\hat{p}^{2})$, taking into account that the intermediate
protons in Fig.~\ref{fig:born_diagrams} are off shell.
This proton off-shell form factor is parametrized here in the exponential form,
\begin{eqnarray}
\hat{F}_{p}(\hat{p}^{2}) = \exp\left( \frac{\hat{p}^{2}-m_{p}^{2}}{\Lambda_{off,E}^{2}} \right) \,,
\label{ff_exp}
\end{eqnarray}
where $\Lambda_{off,E}$ 
has to be adjusted to experimental data.
The form factor (\ref{ff_exp}) 
is normalized to unity at the on-shell point $\hat{p}^{2} = m_{p}^{2}$.

In a way similar to (\ref{amplitude_pompom_t}) - (\ref{amplitude_pompom_approx})
we can write the amplitudes for the exchanges
$(\Pom, \Reg_{+})$, $(\Reg_{+}, \Pom)$, and $(\Reg_{+}, \Reg_{+})$,
since both, $\Pom$ and $\Reg_{+}$ exchange, are treated as tensor exchanges in our model.
The contributions in (\ref{C_modes_ppbar_mm}) - (\ref{C_modes_ppbar_mp})
involving $C = -1$ exchanges are different.
We recall that $\Reg_{-}$ exchanges are treated as effective vector exchanges
in our model; see Sec.~3 of \cite{Ewerz:2013kda} and Appendix~\ref{sec:appendixA}
of the present paper.

\section{$pp \to pp (f_{0} \to p \bar{p})$}
\label{sec:section_f0_ppbar}

The resonances produced diffractively in the $p \bar{p}$ channel are not well known.
Therefore, we will concentrate only 
on the $s$-channel scalar resonances.
We shall study the reaction $pp \to pp (f_{0} \to p \bar{p})$
where $f_{0}$ stands 
for one of the $f_{0}(2020)$, $f_{0}(2100)$, and $f_{0}(2200)$ states
with $I^{G}(J^{PC})= 0^{+}(0^{++})$.
It must be noted that these states are only listed in \cite{Olive:2016xmw}
and are not included in the summary tables.
Also their couplings to the $p \bar{p}$ channel are essentially unknown.

The $\Pom \Pom$-exchange amplitude
through a scalar resonance $f_{0} \to p \bar{p}$
can be written as
\begin{equation}
\begin{split}
& {\cal M}^{(\Pom \Pom \to f_{0} \to p \bar{p})}_{\lambda_{a}\lambda_{b}
\to\lambda_{1}\lambda_{2}\lambda_{3}\lambda_{4}}
\simeq 3 \beta_{\Pom NN}  \, 2(p_1 + p_a)_{\mu_{1}} (p_1 + p_a)_{\nu_{1}}\, 
\delta_{\lambda_{1} \lambda_{a}}\, F_1(t_1)  \;
\frac{1}{4 s_{1}} (- i s_{1} \alpha'_{\Pom})^{\alpha_{\Pom}(t_{1})-1} \\ 
& \quad \quad
\times 
\Gamma^{(\Pom \Pom f_{0})\,\mu_{1} \nu_{1}, \mu_{2} \nu_{2}}(q_{1},q_{2})\,
\Delta^{(f_{0})}(p_{34})\,
\bar{u}(p_{4}, \lambda_{4}) \, \Gamma^{(f_{0} p \bar{p})}(p_{4},-p_{3})\,
v(p_{3}, \lambda_{3}) \\
& \quad \quad\times 
\frac{1}{4 s_{2}} (- i s_{2} \alpha'_{\Pom})^{\alpha_{\Pom}(t_{2})-1}\,
3 \beta_{\Pom NN}  \, 2 (p_2 + p_b)_{\mu_{2}} (p_2 + p_b)_{\nu_{2}}\, 
\delta_{\lambda_{2} \lambda_{b}}\, F_1(t_2) \,,
\end{split}
\label{amplitude_approx_ppbar}
\end{equation}
where
$s_{1} = (p_{1} + p_{3} + p_{4})^{2}$,
$s_{2} = (p_{2} + p_{3} + p_{4})^{2}$, 
and $p_{34} = p_{3} + p_{4}$.

The effective Lagrangians and the vertices for $\Pom \Pom$ fusion into 
an $f_{0}$ meson are discussed in Appendix~A of \cite{Lebiedowicz:2013ika}.
As was shown there the tensorial $\Pom \Pom f_{0}$ vertex 
corresponds to the sum of the two lowest values of $(l,S)$, 
that is, $(l,S) = (0,0)$ and $(2,2)$
with coupling parameters 
$g_{\Pom \Pom f_{0}}'$ and $g_{\Pom \Pom f_{0}}''$, respectively.
The vertex, including a form factor, reads then as follows 
($p_{34} = q_{1} + q_{2}$):
\begin{eqnarray}
i\Gamma_{\mu \nu,\kappa \lambda}^{(\Pom \Pom f_{0})} (q_{1},q_{2}) =
\left( i\Gamma_{\mu \nu,\kappa \lambda}'^{(\Pom \Pom f_{0})}\mid_{bare} +
       i\Gamma_{\mu \nu,\kappa \lambda}''^{(\Pom \Pom f_{0})} (q_{1}, q_{2})\mid_{bare} \right)
\tilde{F}^{(\Pom \Pom f_{0})}(q_{1}^{2},q_{2}^{2},p_{34}^{2}) \,;
\label{vertex_pompomS}
\end{eqnarray}
see Eq.~(A.21) of \cite{Lebiedowicz:2013ika}.
As was shown in \cite{Lebiedowicz:2013ika} these two
$(l,S)$ couplings give different results 
for the distribution in the azimuthal angle 
between the transverse momenta $\vec{p}_{t,1}$ and $\vec{p}_{t,2}$ 
of the outgoing leading protons.
We take the factorized form for the pomeron-pomeron-meson form factor
\begin{eqnarray}
\tilde{F}^{(\Pom \Pom f_{0})}(q_{1}^{2},q_{2}^{2},p_{34}^{2}) = 
F_{M}(q_{1}^{2}) F_{M}(q_{2}^{2}) F^{(\Pom \Pom f_{0})}(p_{34}^{2})\,
\label{Fpompommeson}
\end{eqnarray}
normalised to
$\tilde{F}^{(\Pom \Pom f_{0})}(0,0,m_{f_{0}}^{2}) = 1$.
We will further set
\begin{eqnarray}
F^{(\Pom \Pom f_{0})}(p_{34}^{2}) = 
\exp{ \left( \frac{-(p_{34}^{2}-m_{f_{0}}^{2})^{2}}{\Lambda_{f_{0}}^{4}} \right)}\,,
\quad \Lambda_{f_{0}} = 1\;{\rm GeV}\,.
\label{Fpompommeson_ff}
\end{eqnarray}

The scalar-meson propagator is taken as
\begin{eqnarray}
i\Delta^{(f_{0})}(p_{34}) = \dfrac{i}{p_{34}^{2}-m_{f_{0}}^2+i m_{f_{0}} 
\Gamma_{f_{0}}}\,,
\label{prop_scalar}
\end{eqnarray}
with constant widths for the $f_{0}$ states
with the numerical values from \cite{Olive:2016xmw}.

For the $f_{0} p \bar{p}$ vertex we have 
\begin{eqnarray}
i\Gamma^{(f_{0} p \bar{p})}(p_{4},-p_{3}) = 
i g_{f_{0} p \bar{p}} \, F^{(f_{0} p \bar{p})}(p_{34}^{2})\,,
\end{eqnarray}
where $g_{f_{0} p \bar{p}}$ is an unknown dimensionless parameter.
We assume $g_{f_{0} p \bar{p}}>0$ and 
$F^{(f_{0} p \bar{p})}(p_{34}^{2})$ = $F^{(\Pom \Pom f_{0})}(p_{34}^{2})$;
see Eq.~(\ref{Fpompommeson_ff}).

\section{First results}
\label{sec:results}

We start our analysis by comparing the cross section 
of our non-resonant contribution to 
the $pp \to pp p\bar{p}$ reaction (\ref{2to4_reaction})
with the CERN ISR data at $\sqrt{s} = 62$~GeV \cite{Breakstone:1989ty}.
In \cite{Breakstone:1989ty} the centrally produced
antiproton and proton were restricted to lie in the rapidity regions
$|{\rm y}_{3}|$, $|{\rm y}_{4}| < 1.5$, respectively, 
and the outgoing forward protons to have
$x_{F,p} > 0.9$ and the four-momentum transfer squared
$|t| \geqslant 0.08$~GeV$^{2}$.
With such kinematic conditions
we get the integrated cross section of $\sigma_{th} = 0.013$ and 0.236~$\mu$b 
for $\Lambda_{off,E} = 0.8$ and 1~GeV, respectively, 
compared with $\sigma_{exp} = 0.80 \pm 0.17$~$\mu$b from \cite{Breakstone:1989ty}.
Our theoretical results have been obtained in the Born approximation
(neglecting absorptive corrections).
The realistic cross section can be obtained by multiplying 
the Born cross section by the corresponding
gap survival factor $\langle S^{2} \rangle$.
At the ISR energies 
we estimate it to be $\langle S^{2} \rangle \simeq 0.5$.~\footnote{In 
exclusive reactions, as the $pp \to pp \pi^{+}\pi^{-}$ one, for instance, 
the gap survival factor is strongly dependent on the $t_{1}$
and $t_{2}$ variables; see e.g.~\cite{Lebiedowicz:2014bea,Lebiedowicz:2015eka}.}
In our calculations, we include the pomeron 
and reggeon $\Reg_{+}$ and $\Reg_{-}$ exchanges; 
see (\ref{C_modes_ppbar_pp}) - (\ref{C_modes_ppbar_mp}).
For double pomeron exchange and $\Lambda_{off,E} = 1$~GeV
we get only $\sigma_{th} = 0.077$~$\mu$b.
It is seen that inclusion of subleading reggeon
exchanges is crucial at the ISR energy.
In \cite{Akesson:1985rn} the measurement was performed at $\sqrt{s} = 63$~GeV,
$|{\rm y}_{3}|, |{\rm y}_{4}| \leqslant 1$,
$x_{F,p} > 0.95$, $0.01 \lesssim |t| \lesssim 0.06$~GeV$^{2}$
and the cross section $d^{2}\sigma_{exp}/dt_{1}dt_{2}
= 1.0 \pm 0.5$~$\mu$b~GeV$^{-4}$ 
for $t_{1} = t_{2} = -0.035$~GeV$^{2}$ was determined.
We get (without absorption)
$d^{2}\sigma_{th}/dt_{1}dt_{2} = 0.73$
and 14.14~$\mu$b~GeV$^{-4}$  
for $\Lambda_{off,E} = 0.8$
and 1~GeV, respectively. 
We see that this experiment supports the smaller value of $\Lambda_{off,E}$.
Although the ISR experiments \cite{Akesson:1985rn,Breakstone:1989ty} 
were performed for different kinematic coverage,
in both an enhancement in the low $p \bar{p}$ invariant-mass region was observed.
The low-mass enhancement is clearly seen also at the WA102 energy \cite{Barberis:1998sr},
see Fig.~1~(b) there.
Therefore, the non-resonant (continuum) contribution alone is not sufficient to describe the low-energy data
and, e.g., scalar and/or tensor resonance contributions 
should be taken into account.
We will return to this issue below (see Fig.~\ref{fig:f0}).

Now we show numerical results for the reaction $pp \to pp p\bar{p}$
at higher energies.
In Table~\ref{tab:table} we have collected cross sections in $\mu b$
for the exclusive $p \bar{p}$ continuum
including some experimental cuts.
We show results for the pomeron and reggeon exchanges
in the amplitude (see the column ``$\Pom$ and $\Reg$'')
and when only the $(\Pom, \Pom)$ term contributes
(see the column ``$\Pom$'').
The calculations have been done in the Born approximation (without absorption effects)
and for $\Lambda_{off,E} = 1$~GeV in (\ref{ff_exp}). 
The absorption effects lead to a damping
of the cross section by a factor 5 for $\sqrt{s} = 0.2$~TeV and
by a factor 10 for $\sqrt{s} = 13$~TeV; see e.g.~\cite{Lebiedowicz:2015eka}.
The next-to-last line in Table~\ref{tab:table}
shows result with an extra cut on leading protons
of 0.17~GeV~$< |p_{y,1}|, |p_{y,2}|<$~0.5~GeV 
that will be measured in ALFA on both sides of the ATLAS detector.
\begin{table}
\begin{small}
\caption{The integrated cross sections in $\mu$b 
for the exclusive diffractive $p \bar{p}$ continuum production
for some experimental cuts
on (pseudo)rapidity and $p_{t}$ of
centrally produced individual $p$ and $\bar{p}$
for the STAR, ALICE, ATLAS, CMS, and LHCb experiments.
Results for some limitations on leading protons are also shown.
The column ``$\Pom$ and $\Reg$'' 
shows the resulting total cross sections from 
$\Pom$ and $\Reg$ ($\Reg_{+}$ and $\Reg_{-}$) exchanges,
which include, of course, the interference term between the various components.
The column ``$\Pom$'' shows results obtained for the $\Pom$ exchange alone.
We have taken here $\Lambda_{off,E} = 1$~GeV.
No absorption effects were included here.
}
\label{tab:table}
\begin{center}
\begin{tabular}{|c|l|c|c|c|}
\hline
$\sqrt{s}$, TeV & Cuts &$\Pom$ and $\Reg$  & $\Pom$ \\
\hline
0.2 &$|\eta| < 1$, $p_{t} > 0.2$~GeV     & 0.031  & 0.018 \\ 
0.2 &$|\eta| < 1$, $p_{t} > 0.2$~GeV, $0.03 < -t_{1,2} < 0.3$~GeV$^{2}$     & 0.014  & 0.008 \\ 
13 &$|\eta| < 0.9$, $p_{t} > 0.1$~GeV      & 0.032 & 0.031 \\ 
13 &$|\rm{y}| < 2$, $p_{t} > 0.2$~GeV      & 2.38  & 2.19 \\
13 &$|\eta| < 2.5$, $p_{t} > 0.1$~GeV      & 1.96  & 1.82 \\
13 &$|\eta| < 2.5$, $p_{t} > 0.1$~GeV,  $0.17 < |p_{y}| < 0.5$~GeV      & 0.31 & 0.29 \\
13 &$2 < \eta < 4.5$, $p_{t} > 0.2$~GeV        & 0.79 & 0.68 \\
\hline
\end{tabular}
\end{center}
\end{small}
\end{table}

We have also calculated the corresponding cross sections for 
the $pp \to pp \Lambda \overline{\Lambda}$ reaction,
taking into account only the dominant $(\Pom, \Pom)$ contribution.
The amplitude ${\cal M}^{(\Pom \Pom \to \Lambda \overline{\Lambda})}$ is
very much the same as
${\cal M}^{(\Pom \Pom \to p \bar{p})}$ (\ref{amplitude_pompom_approx}) 
but with $m_{p}$, $\hat{F}_{p}$ replaced by $m_{\Lambda}$, $\hat{F}_{\Lambda}$. 
To describe the off-shellness of the intermediate $t/u$-channel
$\Lambda$ baryons we assume the form factor given by
Eq.~(\ref{ff_exp}) with $\Lambda_{off,E} = 1$~GeV.
For the coupling of the $\Lambda$ baryon to the pomeron
we make an ansatz similar to the proton-pomeron coupling
in (\ref{A4}) of Appendix~\ref{sec:appendixA} but with $\beta_{\Pom NN}$
replaced by a constant $\beta_{\Pom \Lambda \Lambda}$.
The value of the latter can be estimated from the data
on the total cross sections for $\Lambda p$ and $pp$ scattering
at high energies 
\footnote{
The $\Lambda p$ total cross sections were measured in Refs.~\cite{Alexander:1969cx,Bassano:1967kbh,Kadyk:1971tc,Gjesdal:1972zu}.
In Ref.~\cite{Gjesdal:1972zu} the average cross section was obtained as
$\sigma_{tot}(\Lambda p) = 34.6 \pm 0.4$~mb
in the hyperon momentum interval $P_{lab} = 6 - 21$~GeV
(which corresponds to $\sqrt{s} \sim 4 - 6$~GeV).
The lack of $\sigma_{tot}(\Lambda p)$ data 
at higher energy does not allow any reasonable estimate of the ratio,
$\sigma_{tot}(\Lambda p)/\sigma_{tot}(pp)$,
for the pomeron part alone.
Instead we can argue that this ratio should be less than 1,
similar to $\sigma_{tot}^{(\Pom)}(K^{+} p)/\sigma_{tot}^{(\Pom)}(\pi^{+}p) < 1$;
see Sec. 3.1 of Ref.~\cite{Donnachie:2002en}.
The factor 0.9 in Eq.~(\ref{beta_PLL}) is our educated guess.
Data files and plots of various hadronic cross sections
can be found in Ref.~\cite{PDG_CrossSections}.}
using (6.41) of \cite{Ewerz:2013kda}
\begin{eqnarray}
\beta_{\Pom \Lambda \Lambda} \cong \beta_{\Pom NN} 
\frac{\sigma_{tot}(\Lambda p)}{\sigma_{tot}(pp)} \cong 1.87\;{\rm GeV}^{-1} \times 0.9 \simeq 1.68\;{\rm GeV}^{-1}\,.
\label{beta_PLL}
\end{eqnarray}
We find a cross section of 0.11~$\mu$b for $\sqrt{s} = 13$~TeV 
and the ATLAS cuts ($|\eta| < 2.5$, $p_{t} > 0.1$~GeV 
on centrally produced $\Lambda$ and $\overline{\Lambda}$ baryons)
and a cross section of 0.04~$\mu$b for the LHCb cuts ($2 < \eta < 4.5$ and $p_{t} > 0.2$~GeV).
The calculated cross section 
for the $\Lambda \overline{\Lambda}$ continuum production
is about 16 times smaller than the corresponding $p \bar{p}$ continuum production one.

\begin{figure}[!ht]
\includegraphics[width=0.48\textwidth]{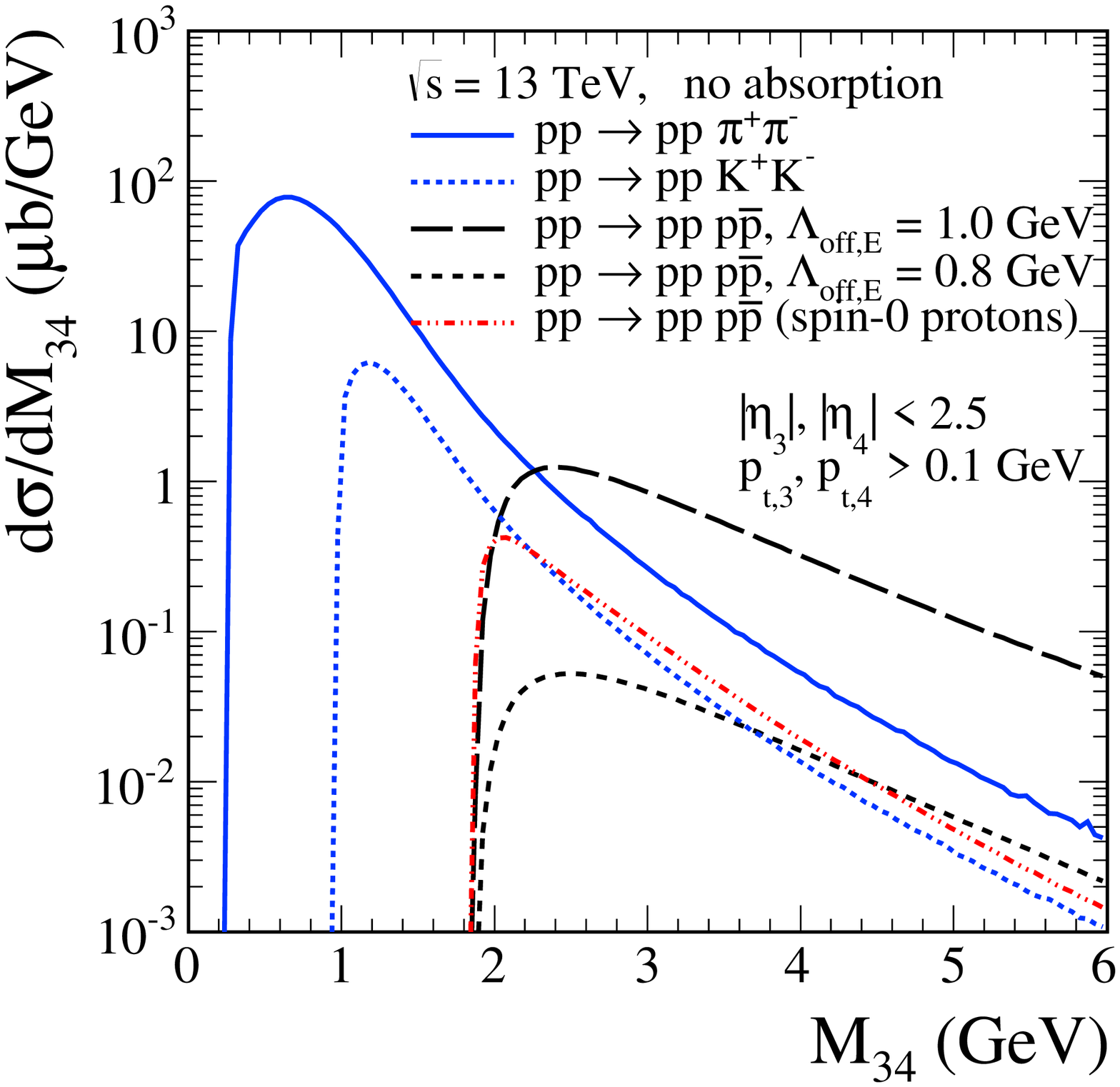}
\includegraphics[width=0.48\textwidth]{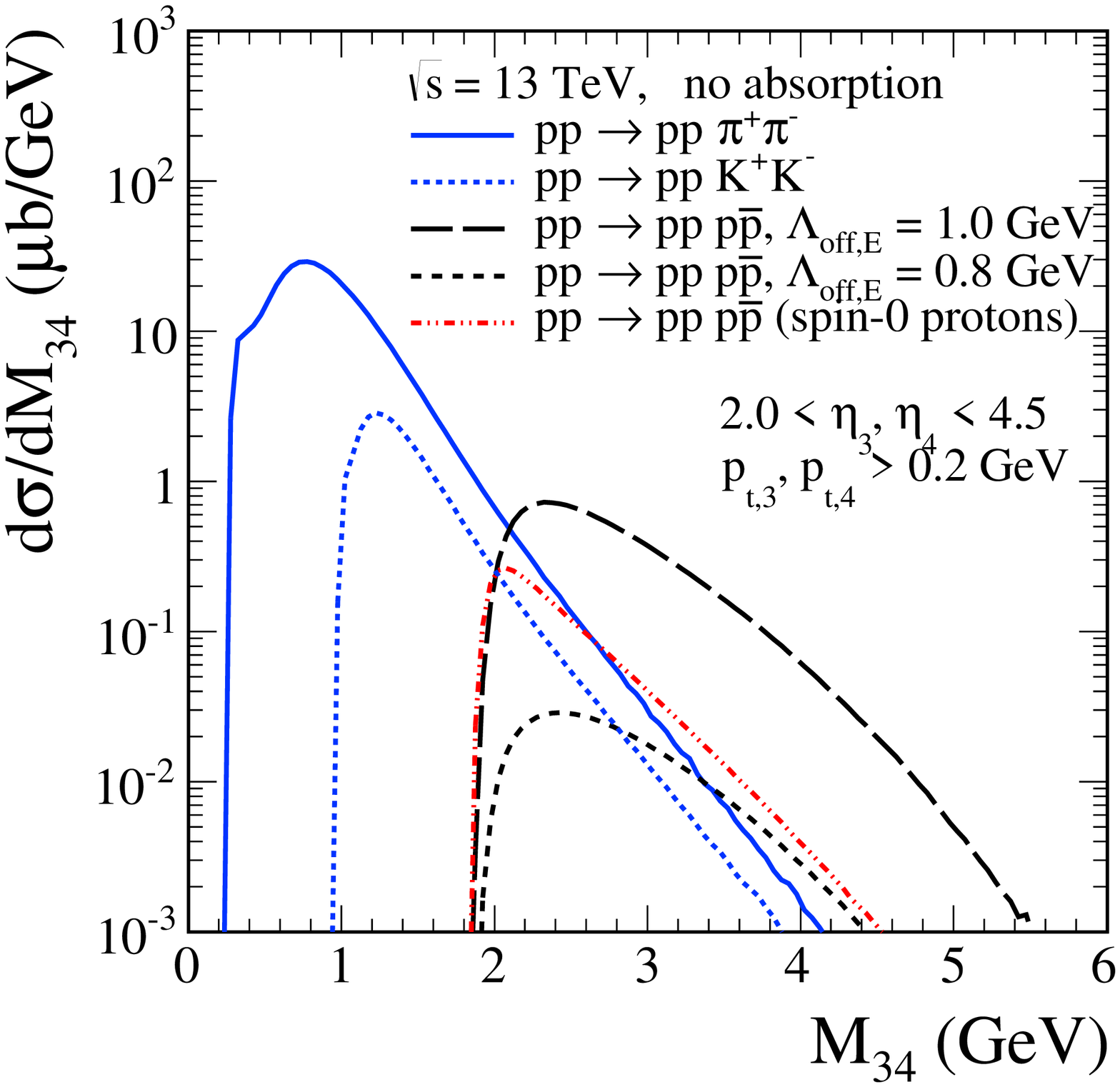}
  \caption{\label{fig:dsig_dMhh}
  \small
The invariant mass distributions for 
centrally produced $\pi^+ \pi^-$, $K^+ K^-$ and $p \bar{p}$ systems
for different experimental conditions at $\sqrt{s} = 13$~TeV.
Results for the combined tensor-pomeron and reggeon exchanges
and $\Lambda_{off,E} = 1$~GeV are presented.
For the $p \bar{p}$ production we show results 
also for $\Lambda_{off,E} = 0.8$~GeV; see (\ref{ff_exp}).
No absorption effects were included here.}
\end{figure}

\begin{figure}[!ht]
\includegraphics[width=0.48\textwidth]{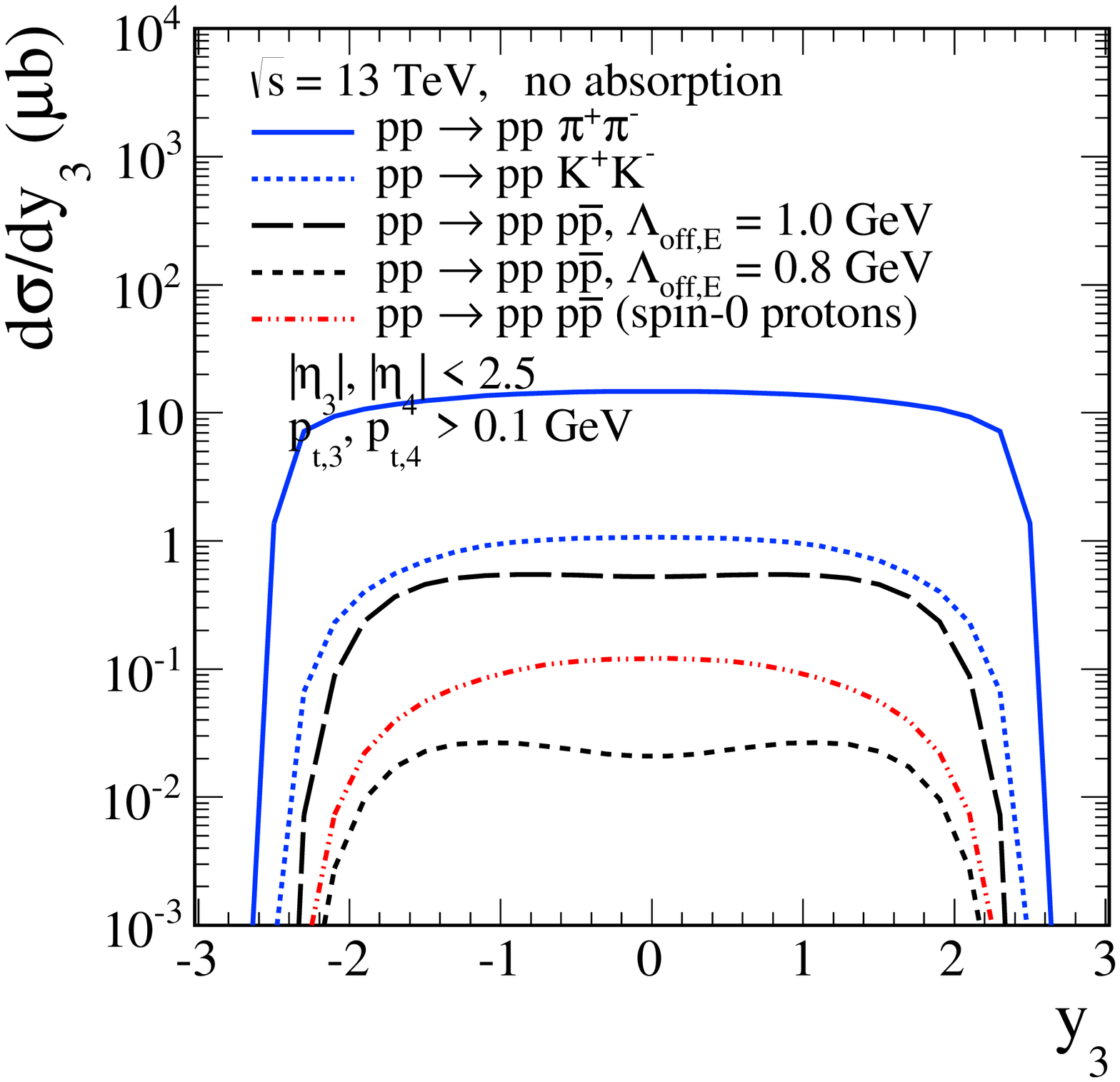}
\includegraphics[width=0.48\textwidth]{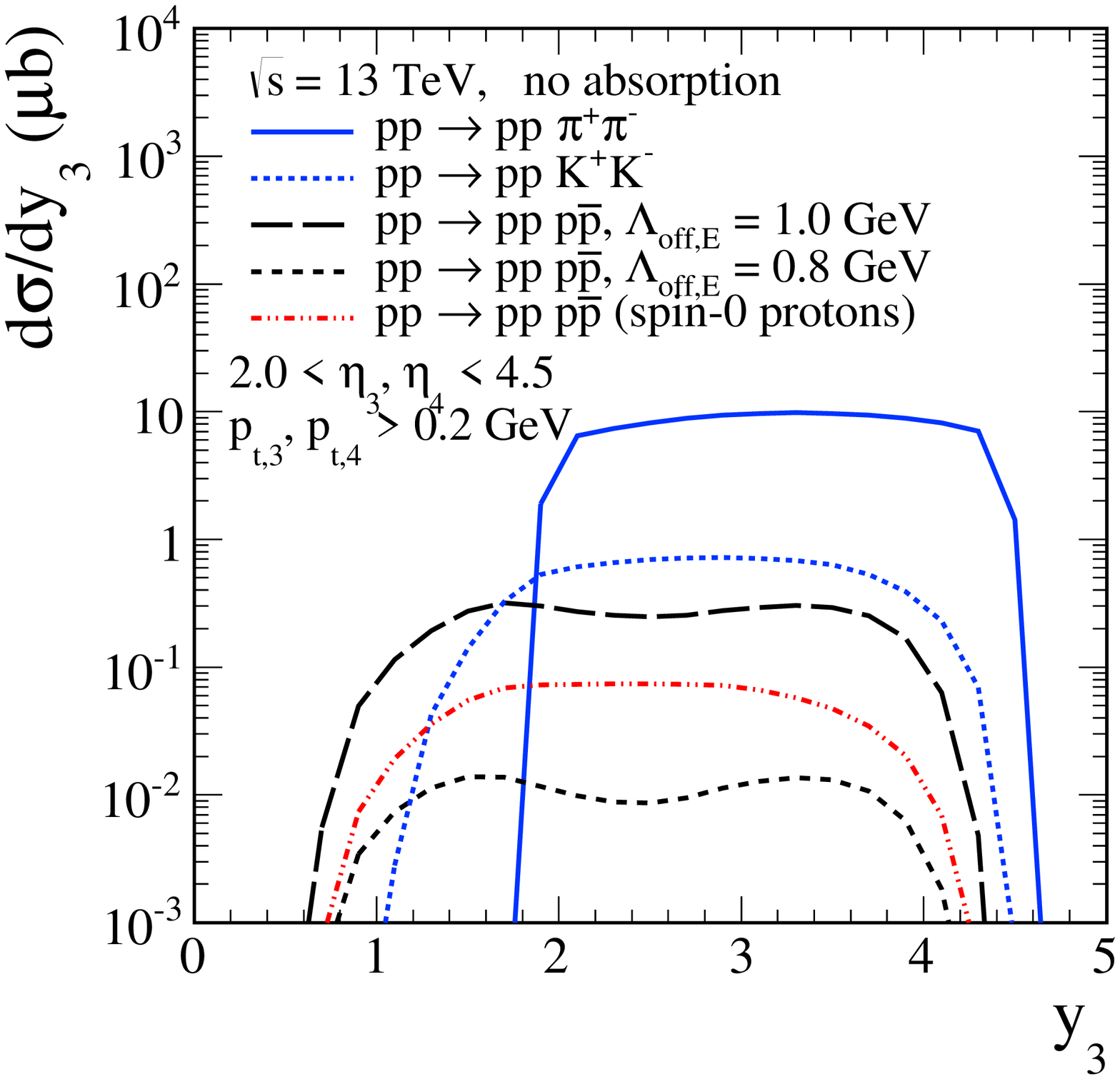}
\includegraphics[width=0.48\textwidth]{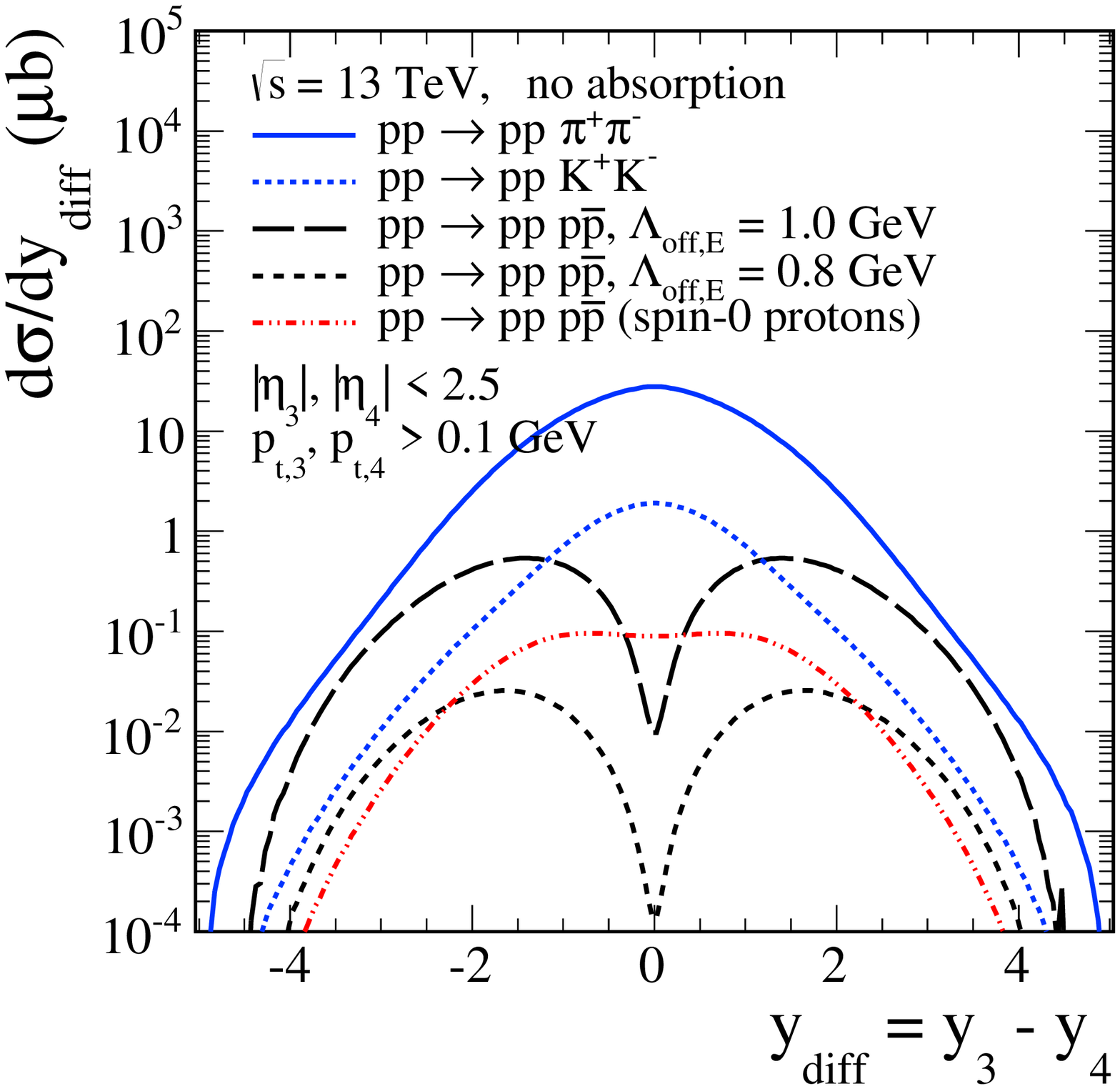}
\includegraphics[width=0.48\textwidth]{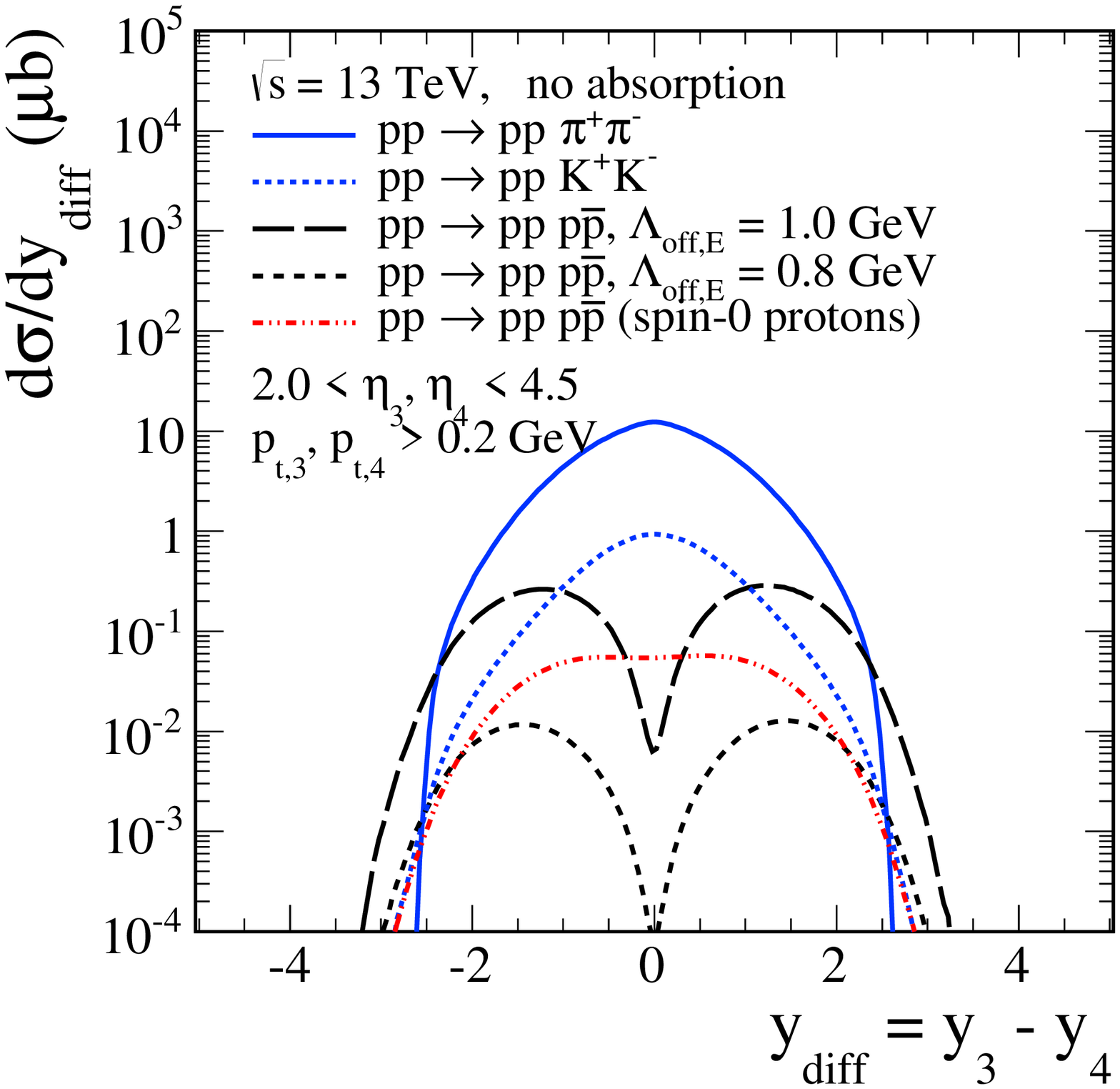}
  \caption{\label{fig:dsig_dy}
  \small
The differential cross sections for various processes at $\sqrt{s} = 13$~TeV.
In the top panels we show the rapidity distributions obtained 
for the tensor pomeron and reggeon exchanges.
In the bottom panels we show the distributions in the rapidity difference 
between the centrally produced hadrons.
No absorption effects were included here.}
\end{figure}

In Figs.~\ref{fig:dsig_dMhh} and \ref{fig:dsig_dy},
we present the distributions in the $p \bar{p}$ invariant mass $M_{34}$,
in the antiproton rapidity ${\rm y}_{3}$, 
and in the rapidity distance between the antiproton and proton 
${\rm y}_{diff} = {\rm y}_{3}-{\rm y}_{4}$ at $\sqrt{s} = 13$~TeV.
We wanted to concentrate only on the main characteristics 
of the $p \bar{p}$ continuum production;
therefore, the calculations have been done neglecting the absorptive corrections.
To illustrate uncertainties of our model, we take in the calculation
two values of $\Lambda_{off,E}$; see Eq.~(\ref{ff_exp}).
The black long-dashed line represents the result for $\Lambda_{off,E} = 1$~GeV,
and the black short-dashed line represents the result for $\Lambda_{off,E} = 0.8$~GeV.
For comparison, we also show results for the $\pi^+ \pi^-$ and $K^+ K^-$ continuum production,
see the blue solid line and the blue dotted line, respectively.
The reaction $pp \to pp \pi^{+} \pi^{-}$ was discussed already in~\cite{Lebiedowicz:2016ioh}.
The reaction $pp \to pp K^{+} K^{-}$ in the tensor-pomeron approach
was recently studied in~\cite{Lebiedowicz:2018eui}.
For reference, we show also a naive (``spin-0 protons'') result 
for artificially modified spin of centrally produced nucleons, from 1/2 to 0;
see the red dash-dotted line.
Here, we assume the amplitude as for $K^{+} K^{-}$ production 
but with some modifications,
e.g., in the case of the $(\Pom,\Pom)$ term 
replacing $m_{K}$, $2 \beta_{\Pom K K}$, and $F_{M}(t_{1,2})$  
by $m_{p}$, $3 \beta_{\Pom NN}$, and $F_{1}(t_{1,2})$, respectively.
We take into account also reggeon exchanges with the corresponding
reggeon-nucleon-nucleon coupling parameters.
This result is purely academic but illustrates how important
the correct inclusion of the spin degrees of freedom is in the Regge calculation.
Different spin of the produced particles clearly leads to different results.

In Fig.~\ref{fig:dsig_dMhh} we compare the invariant mass distributions 
for the $\pi^+ \pi^-$, $K^+ K^-$ and $p \bar{p}$ cases
for two different experimental conditions at $\sqrt{s} = 13$~TeV.
In our calculations we have included both pomeron and reggeon exchanges. 
The distribution in $p \bar{p}$ invariant mass has much larger threshold 
but is also much less steep, compared to that for production 
of pseudoscalar meson pairs.
This effect is related to the spin of the produced particles (1/2 versus 0). 
We hope for a confirmation of the slope of the invariant mass distribution,
e.g., by the ATLAS or the ALICE collaboration.
We see from Fig.~\ref{fig:dsig_dMhh} that the normalisations of the $M_{34}$ distributions
for $p \bar{p}$ are very sensitive to the cutoff parameter
$\Lambda_{off,E}$ of (\ref{ff_exp}).
In addition, we have the effects of absorption corrections.
To fix the magnitudes of these two effects, we will have,
at the moment, to have recourse to experimental input which,
presumably, will come soon.

In Fig.~\ref{fig:dsig_dy}, we show the rapidity distributions (the top panels) 
and the distributions in rapidity difference 
${\rm y}_{diff} = {\rm y}_{3} - {\rm y}_{4}$ (the bottom panels)
for the ATLAS and LHCb pseudorapidity ranges.
The distribution in the (anti)proton rapidity looks rather standard
while the distribution for ${\rm y}_{diff}$ is very special.
We predict a dip in the rapidity difference between 
the antiproton and proton for ${\rm y}_{diff} = 0$.
The dip is caused by a good separation of ${\rm \hat{t}}$ and ${\rm \hat{u}}$ 
contributions in (${\rm y}_{3}, {\rm y}_{4}$) space.
This novel effect is inherently related 
to the spin 1/2 of the produced hadrons.
We have checked that for the $p \bar{p}$ production
the ${\rm \hat{t}}$- and ${\rm \hat{u}}$-channel
diagrams interfere destructively for
$(C_{1},C_{2}) = (1,1)$ and $(-1,-1)$ exchanges
and constructively for $(1,-1)$ and $(-1,1)$ exchanges.
For the $\pi^{+} \pi^{-}$ production,
we get the opposite interference effects between
the ${\rm \hat{t}}$- and ${\rm \hat{u}}$-channel diagrams.

\begin{figure}[!ht]
\includegraphics[width=0.48\textwidth]{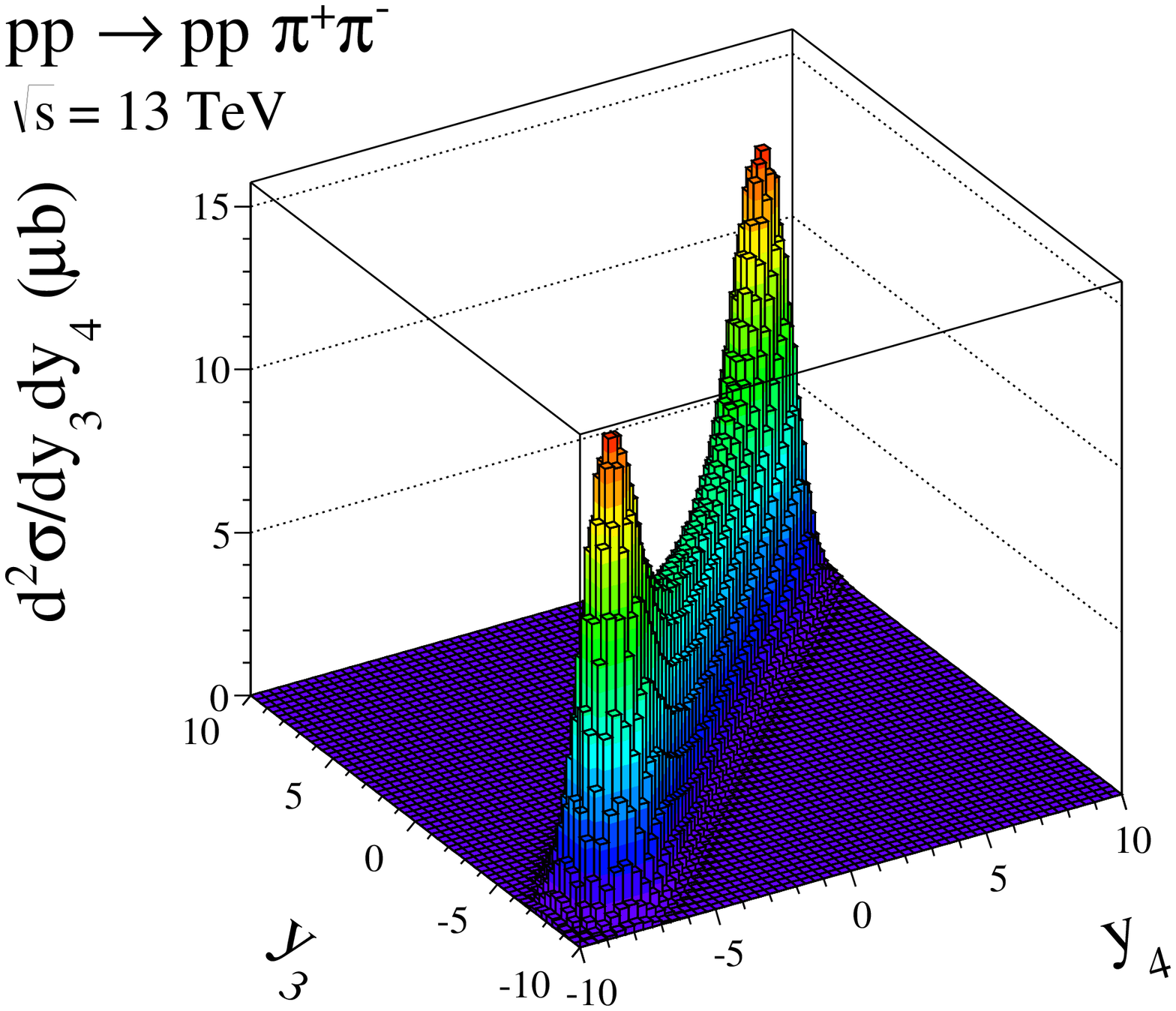}
\includegraphics[width=0.48\textwidth]{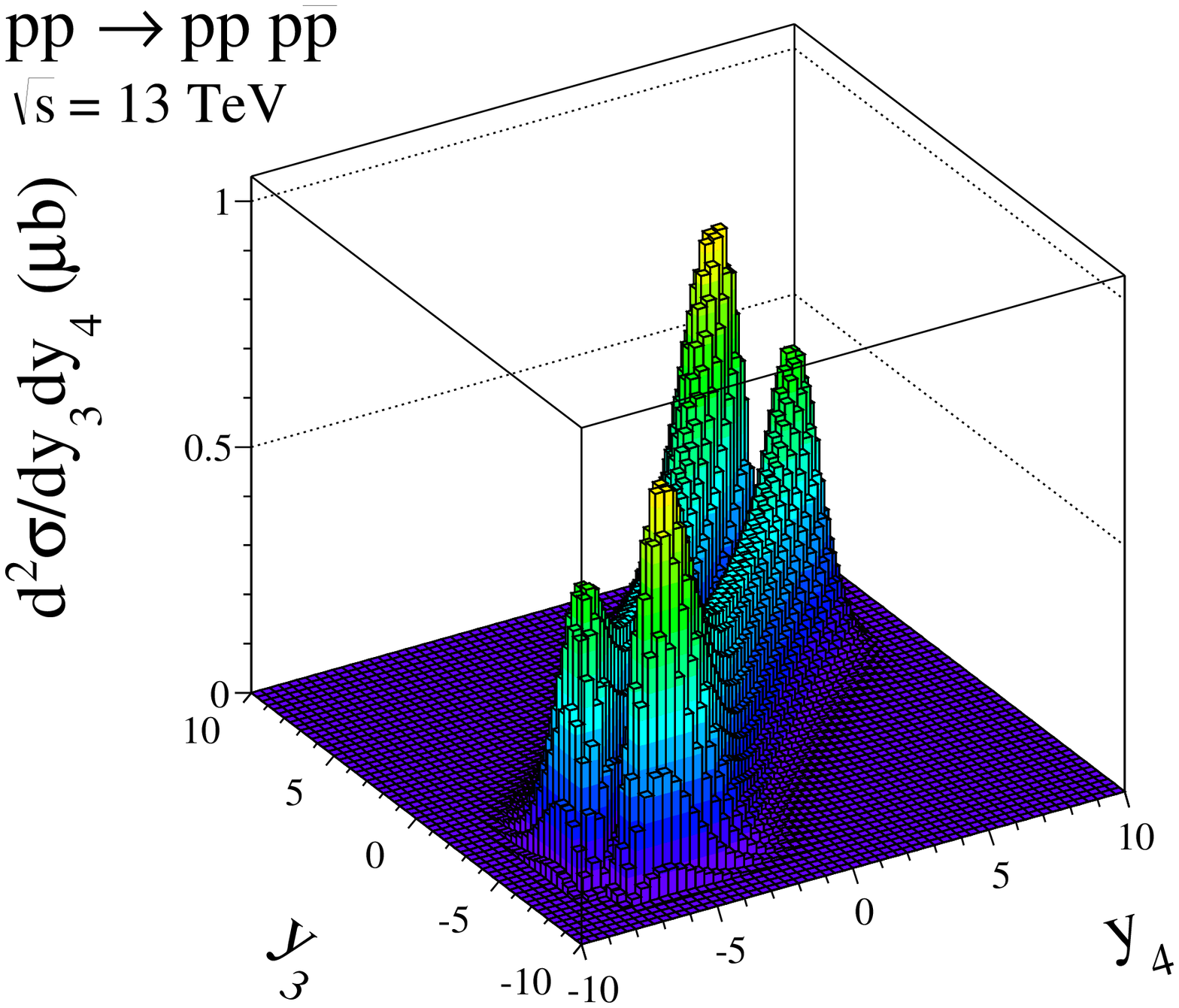}
  \caption{\label{fig:dsig_dy3dy4}
  \small
The two-dimensional distributions in (${\rm y}_{3}, {\rm y}_{4}$) 
for two processes at $\sqrt{s} = 13$~TeV.
Results for the combined tensor-pomeron and reggeon exchanges are presented.
We have taken here $\Lambda_{off,E} = 1$~GeV; see (\ref{ff_exp}).
No absorption effects were included here.
The asymmetry in the right panel will be discussed below.}
\end{figure}
In Fig.~\ref{fig:dsig_dy3dy4} we show the two-dimensional distributions
in rapidity of the $\pi^+$ and $\pi^-$ (the left panel) and
of the antiproton and proton (the right panel) for the full phase space.
In our calculations we have included both pomeron and reggeon exchanges. 
The reggeon exchange contributions lead to enhancements of the cross section
mostly at large rapidities of the centrally produced hadrons.
For the production of the dipion continuum, the cross section 
is concentrated along the diagonal $\rm{y}_{3} = \rm{y}_{4}$.
For the production of $p \bar{p}$ pairs,
one can observe that the dip extends over the whole
diagonal in (${\rm y}_{3}, {\rm y}_{4}$) space.

\begin{figure}[!ht]
\includegraphics[width=0.48\textwidth]{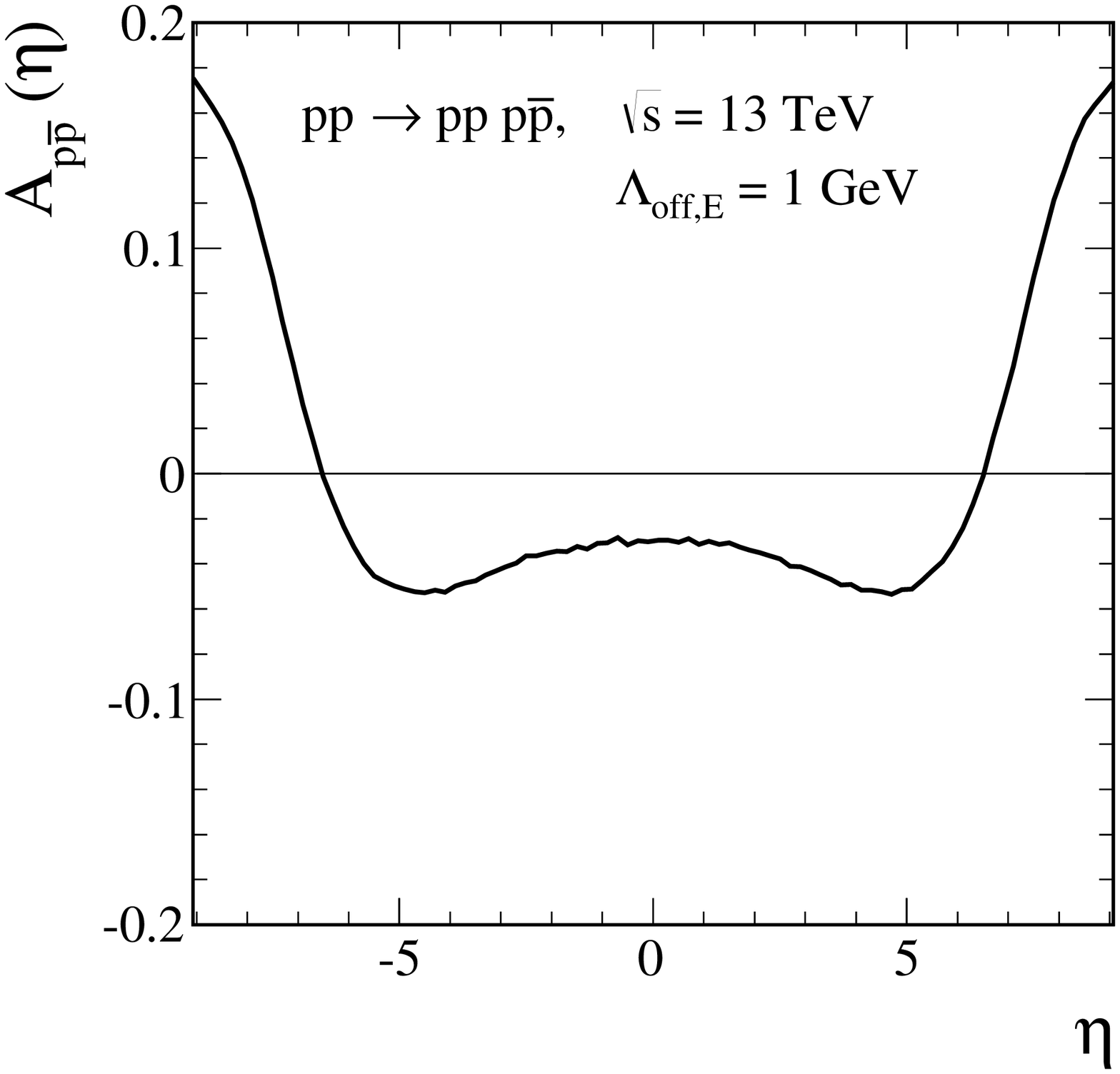}\\
\includegraphics[width=0.48\textwidth]{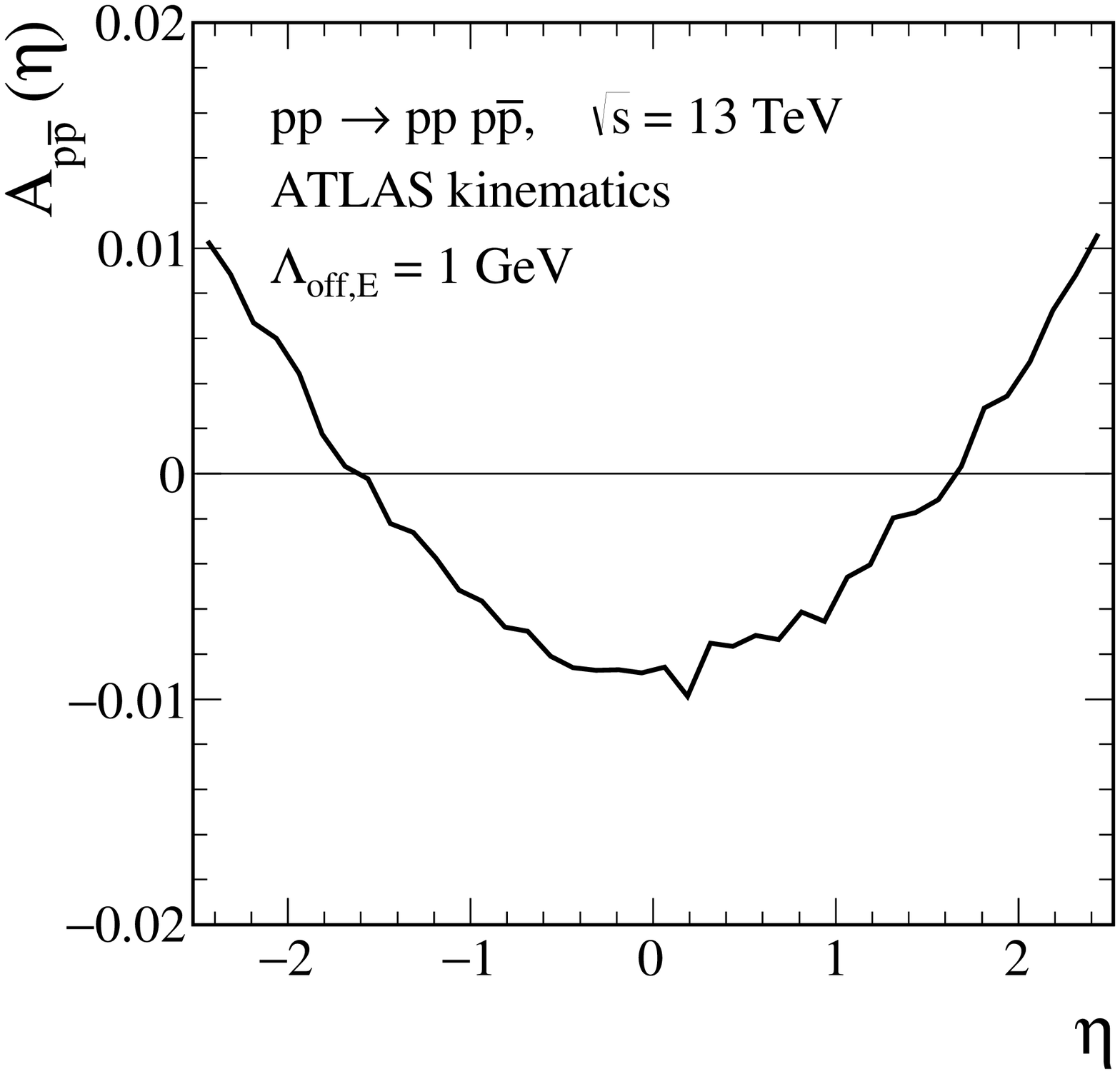}
\includegraphics[width=0.48\textwidth]{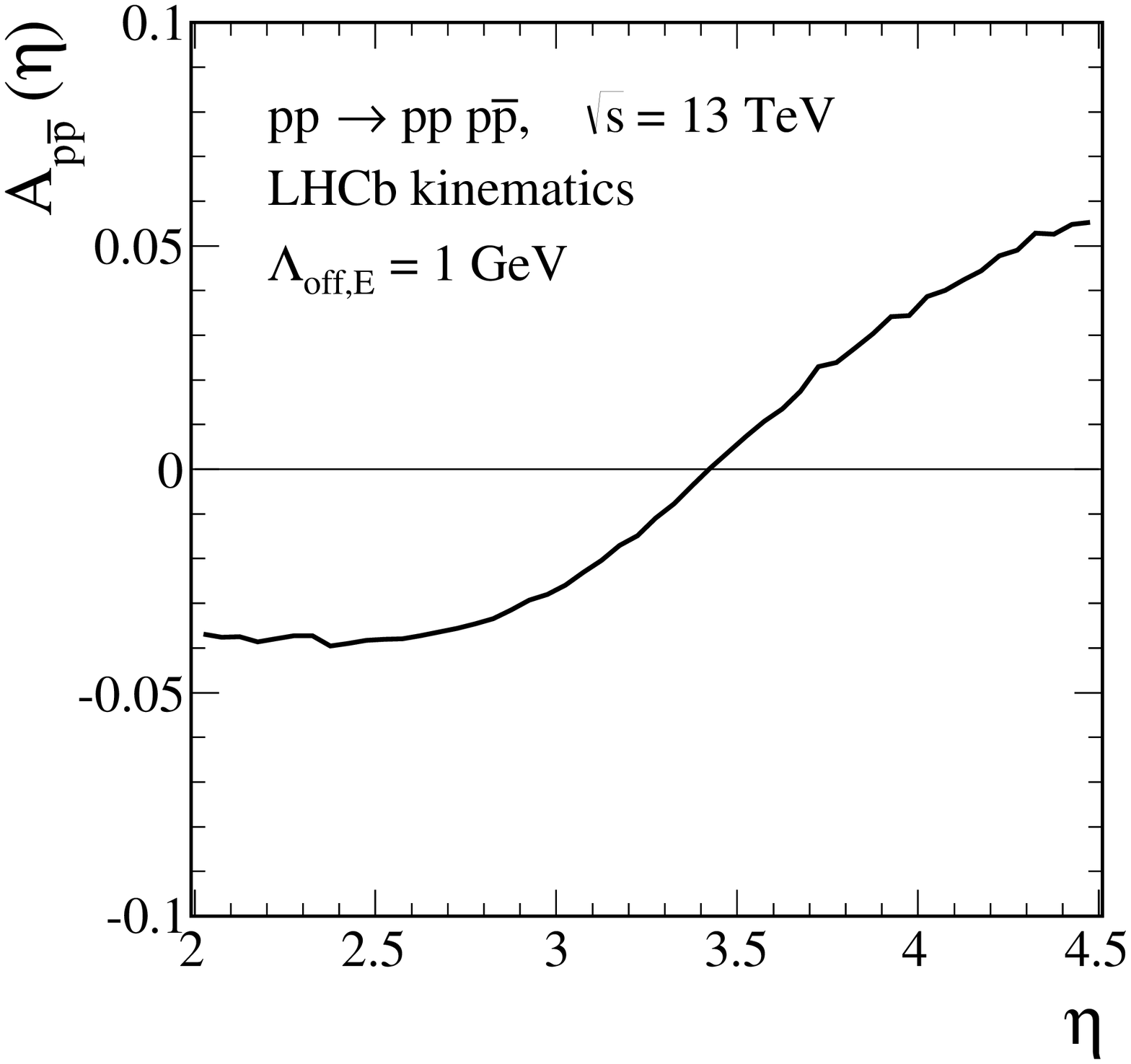}
  \caption{\label{fig:asym}
  \small
The asymmetry $A_{p\bar{p}}(\eta)$ (\ref{asymmetry_eta})
as function of the pseudorapidity $\eta$
for the full phase space (the top panel) and
for the ATLAS and LHCb pseudorapidity ranges (the bottom panels) at $\sqrt{s} = 13$~TeV.
Cuts on the transverse momenta of the centrally produced nucleons
$p_{t,3}, p_{t,4} > 0.1$ and 0.2~GeV for the ATLAS and LHCb, respectively, have been imposed.}
\end{figure}
Figure~\ref{fig:asym} shows the asymmetry
\begin{eqnarray}
A_{p\bar{p}}(\eta) = \frac{\frac{d\sigma}{d\eta_{3}}(\eta) - \frac{d\sigma}{d\eta_{4}}(\eta)}
                          {\frac{d\sigma}{d\eta_{3}}(\eta) + \frac{d\sigma}{d\eta_{4}}(\eta)}\,,
\label{asymmetry_eta}
\end{eqnarray}
where $\eta_{3}$ and $\eta_{4}$ are the pseudorapidities of the antiproton and proton,
respectively,
as a function of the pseudorapidity $\eta$ at $\sqrt{s} = 13$~TeV. 
No absorption effects are included here, but
they should approximately cancel in the ratio.
Sizeable asymmetries are predicted in the full phase space.
Much smaller asymmetries are seen for the limited range 
of pseudorapidities corresponding to the ATLAS, CMS, and ALICE experiments.
The effect is better seen for the LHCb experiment,
which covers the higher pseudorapidity region relevant for the reggeon exchanges.
The asymmetry is caused by the interference of the $(C_{1},C_{2}) = (1,1)$
and $(-1,-1)$ exchanges with the $(C_{1},C_{2}) = (1,-1)$ and $(-1,1)$
exchanges; see (\ref{C_modes_ppbar_pp}) - (\ref{C_modes_ppbar_mp}).
The former exchanges give an amplitude that is antisymmetric under
$p_{3} \leftrightarrow p_{4}$, whereas the latter exchanges
give a symmetric amplitude under $p_{3} \leftrightarrow p_{4}$; 
see (\ref{Tamplitude_aux}) and the discussion following it.
Thus, the resulting $p \bar{p}$ distribution
will not be symmetric under $p_{3} \leftrightarrow p_{4}$.
The biggest contributions to the asymmetry come from
the interference of the $(\Pom,\Pom)$ term
with the $(\Pom, \omega_{\Reg})$ and $(\omega_{\Reg},\Pom)$ contributions
to the total amplitude.
The prediction is that at larger $|\eta|$
more $\bar{p}$ than $p$ should be observed, while
at smaller $|\eta|$, the situation is reversed.

More general asymmetries than (\ref{asymmetry_eta}) can be
considered and are again due to the interference of the $(C_{1},C_{2}) = (1,1)$
and $(-1,-1)$ with the $(C_{1},C_{2}) = (1,-1)$ and $(-1,1)$ exchanges.
We emphasize that the following discussion holds for both
non-resonant and resonant $p \bar{p}$ production.
We can, for instance, consider the one-particle distributions
for the central $p$ and $\bar{p}$ in the overall c.m. system,
\begin{eqnarray}
&&\frac{d^{3}\sigma}{d^{3}p_{3}}(\vec{p}_{3}) \quad {\rm for \;the \;antiproton}\,, \nonumber\\
&&\frac{d^{3}\sigma}{d^{3}p_{4}}(\vec{p}_{4}) \quad {\rm for \;the \;proton}\,, \nonumber
\end{eqnarray}
and the asymmetry
\begin{eqnarray}
A^{(1)}(\vec{p})=\frac{\frac{d^{3}\sigma}{d^{3}p_{3}}(\vec{p})-\frac{d^{3}\sigma}{d^{3}p_{4}}(\vec{p})}
                      {\frac{d^{3}\sigma}{d^{3}p_{3}}(\vec{p})+\frac{d^{3}\sigma}{d^{3}p_{4}}(\vec{p})}\,.
\label{asymmetry_1}
\end{eqnarray}
Here, the leading protons $p(\vec{p}_{1})$ and $p(\vec{p}_{2})$ may be
integrated over their whole or only a part of their phase space.
We can also consider the two-particle cross section for the centrally
produced $p$ and $\bar{p}$: $\frac{d^{6}\sigma}{d^{3}p_{3} d^{3}p_{4}}(\vec{p}_{3},\vec{p}_{4})$.
A suitable asymmetry there is
\begin{eqnarray}
A^{(2)}(\vec{p},\vec{p}')=\frac{\frac{d^{6}\sigma}{d^{3}p_{3} d^{3}p_{4}}(\vec{p},\vec{p}')
                               -\frac{d^{6}\sigma}{d^{3}p_{3} d^{3}p_{4}}(\vec{p}',\vec{p})}
                               {\frac{d^{6}\sigma}{d^{3}p_{3} d^{3}p_{4}}(\vec{p},\vec{p}')
                               +\frac{d^{6}\sigma}{d^{3}p_{3} d^{3}p_{4}}(\vec{p}',\vec{p})}\,.
\label{asymmetry_2}
\end{eqnarray}
In words, this asymmetry means the following.
We choose two momenta $\vec{p}$ and $\vec{p}'$.
Then, we ask if the situations ($\bar{p}(\vec{p}),p(\vec{p}')$)
and ($\bar{p}(\vec{p}'),p(\vec{p})$) occur at the same
or at a different rate.

\begin{figure}[!ht]
\includegraphics[width=0.49\textwidth]{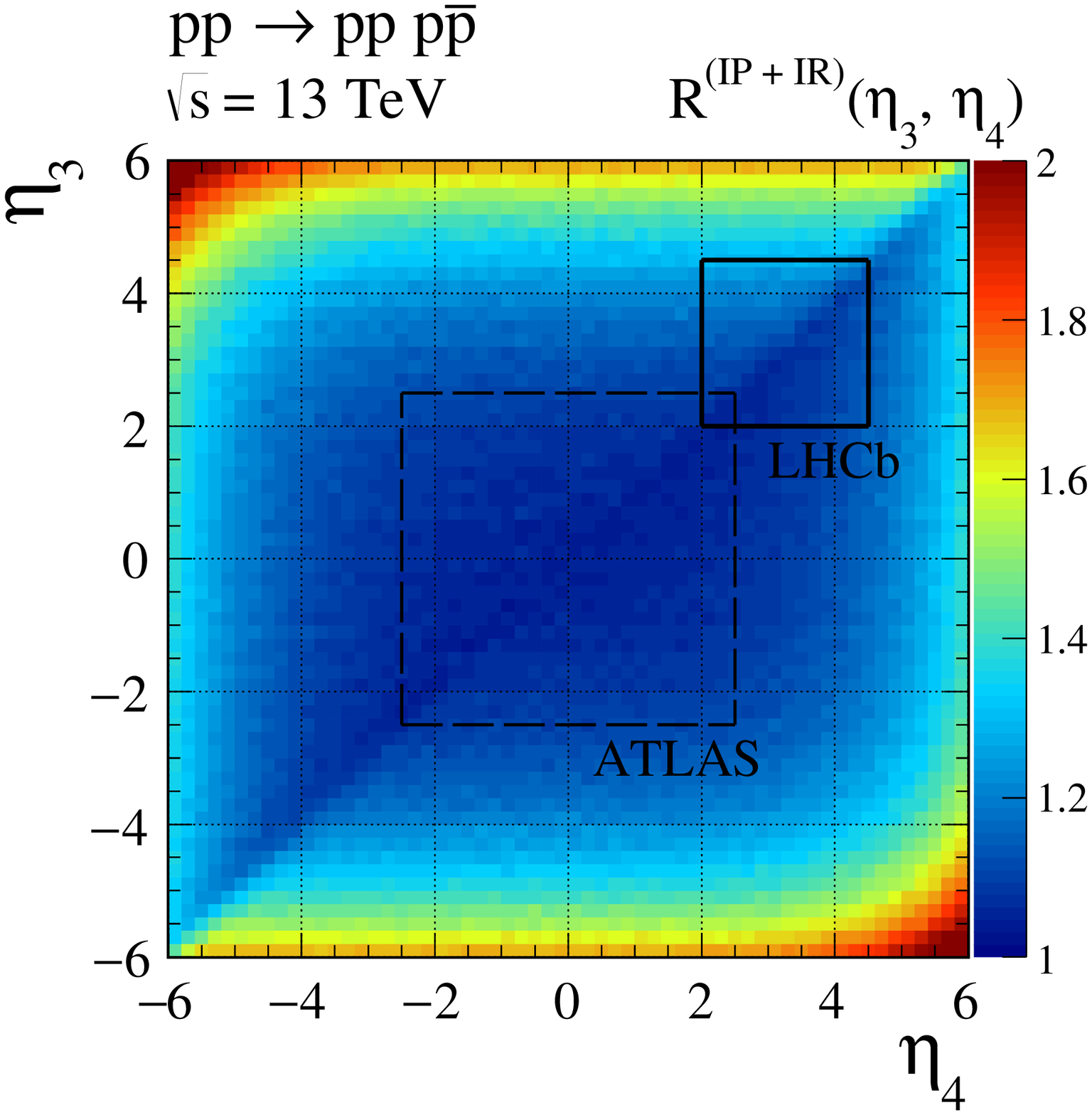}
\includegraphics[width=0.49\textwidth]{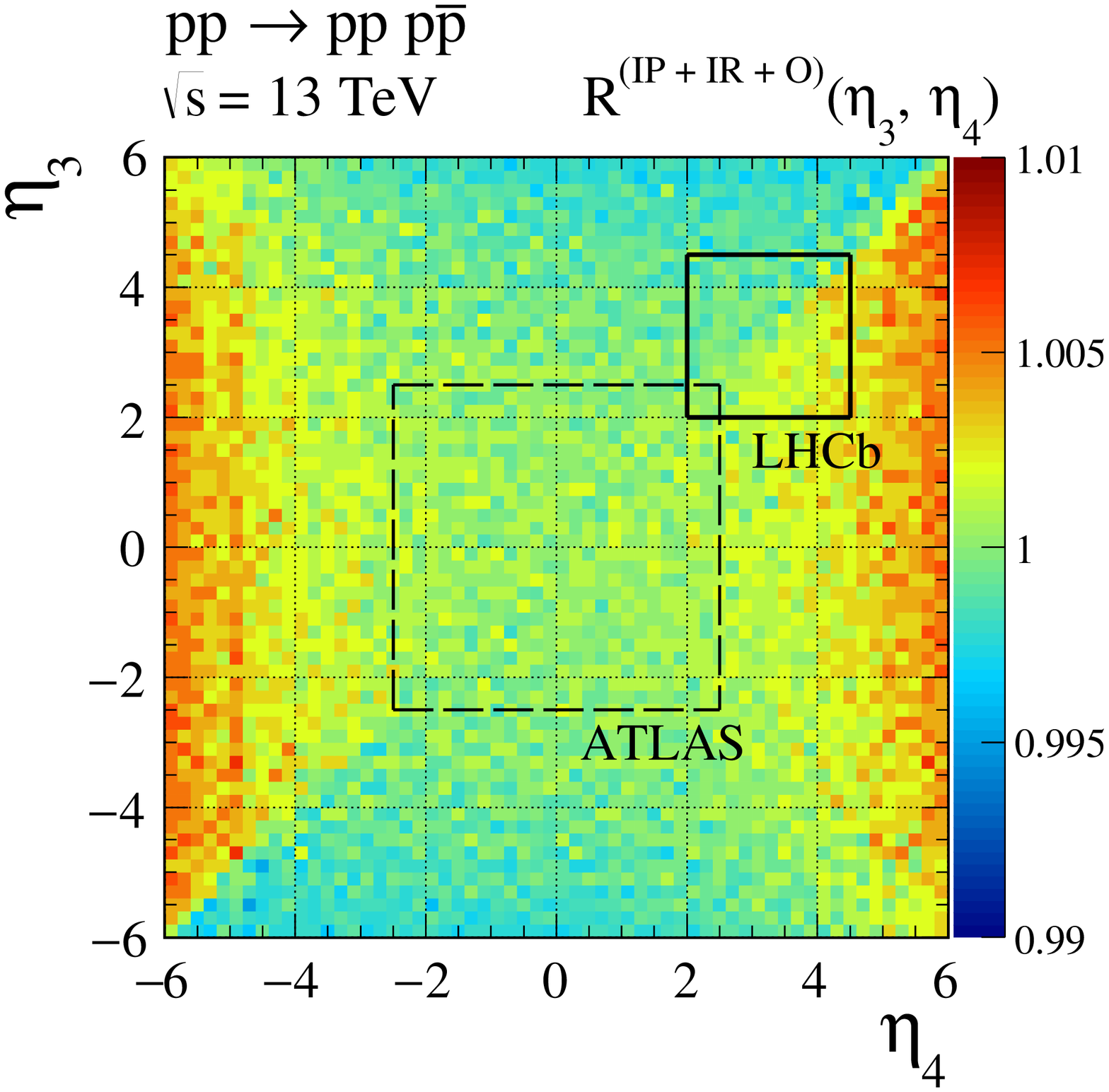}
\caption{\label{fig:5a}
\small
The ratios $R(\eta_{3},\eta_{4})$ at $\sqrt{s} = 13$~TeV and $p_{t,3}, p_{t,4} > 0.2$~GeV.
The left panel represents the result for the ratio 
$R^{(\Pom + \Reg)}(\eta_{3},\eta_{4})$ (\ref{ratio_eta3eta4_PR}).
The right panel shows the result for
$R^{(\Pom + \Reg + \Ode)}(\eta_{3},\eta_{4})$ (\ref{ratio_eta3eta4_PRO}).
Note that different z scales are taken for the left and right panels.
Calculations were done with the parameters of Appendix~A.
In addition regions of the coverage for the ATLAS and LHCb experiments are shown.
}
\end{figure}

Another asymmetry of this type can be constructed
from the pseudorapidity distributions
$\frac{d^{2}\sigma}{d\eta_{3}d\eta_{4}}(\eta_{3},\eta_{4})$.
For two pseudorapidities $\eta$ and $\eta'$, we define
\begin{eqnarray}
\widetilde{A}^{(2)}(\eta,\eta')=\frac{\frac{d^{2}\sigma}{d\eta_{3} d\eta_{4}}(\eta,\eta')
                                 -\frac{d^{2}\sigma}{d\eta_{3} d\eta_{4}}(\eta',\eta)}
                                 {\frac{d^{2}\sigma}{d\eta_{3} d\eta_{4}}(\eta,\eta')
                                 +\frac{d^{2}\sigma}{d\eta_{3} d\eta_{4}}(\eta',\eta)}\,.
\label{asymmetry_3}
\end{eqnarray}
For the quantity 
$\frac{d^{2}\sigma}{d\eta_{3}d\eta_{4}}(\eta_{3},\eta_{4})$
and the asymmetry $\widetilde{A}^{(2)}(\eta,\eta')$,
we have also investigated effects of an odderon 
using the parameters of (\ref{A12}) - (\ref{A14}).
In Fig.~\ref{fig:5a}, we show, in two-dimensional plots,
the ratios
\begin{eqnarray}
&&R^{(\Pom + \Reg)}(\eta_{3},\eta_{4}) = 
\frac{d^{2}\sigma^{(\Pom + \Reg)}/d\eta_{3}d\eta_{4}}{d^{2}\sigma^{(\Pom)}/d\eta_{3}d\eta_{4}}\,,
\label{ratio_eta3eta4_PR}\\
&&R^{(\Pom + \Reg + \Ode)}(\eta_{3},\eta_{4}) = 
\frac{d^{2}\sigma^{(\Pom + \Reg + \Ode)}/d\eta_{3}d\eta_{4}}{d^{2}\sigma^{(\Pom + \Reg)}/d\eta_{3}d\eta_{4}}
\label{ratio_eta3eta4_PRO}
\end{eqnarray}
for $\sqrt{s} = 13$~TeV and $-6 \leqslant \eta_{3}, \eta_{4} \leqslant 6$.
We see that in the limited range of pseudorapidities corresponding to the ATLAS and LHCb experiments
the effects of the secondary reggeons are predicted to be 
in the ranges of 2 - 11~\% and 5 - 26~\%, respectively.
The addition of an odderon with the parameters of (\ref{A14})
has only an effect of less than 0.5~\%.

\begin{figure}[!ht]
\includegraphics[width=0.55\textwidth]{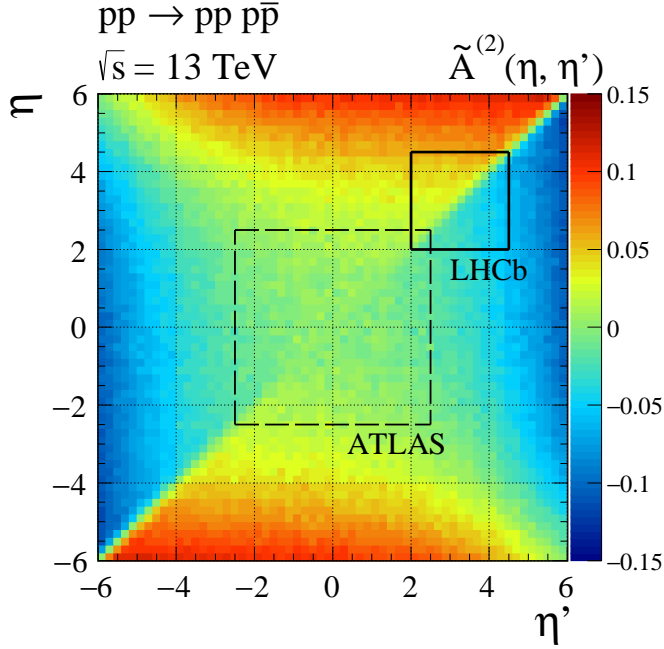}
\caption{\label{fig:5b}
\small
The asymmetry $\widetilde{A}^{(2)}(\eta,\eta')$ (\ref{asymmetry_3})
at $\sqrt{s} = 13$~TeV and $p_{t,3}, p_{t,4} > 0.2$~GeV. 
Shown is the result including pomeron and reggeon exchanges.
}
\end{figure}

In Fig.~\ref{fig:5b} we show the asymmetry (\ref{asymmetry_3})
including pomeron and reggeon exchanges.
For the investigated pseudorapidity range the asymmetries 
due to pomeron plus reggeon exchange show a characteristic pattern:
positive for $|\eta| > |\eta'|$ and negative for $|\eta| < |\eta'|$.
That is, antiprotons are predicted to come out typically
with a higher absolute value of the (pseudo)rapidity than protons.
In Fig.~\ref{fig:5b} the inclusion of the odderon would hardly
change the result, only at the level of less than 1~\%.
This is less than theoretical uncertainties associated 
with the reggeon exchanges.

Finally, we note that for calculations of the asymmetries 
(\ref{asymmetry_eta}) - (\ref{asymmetry_3})
it is essential to use a model in which the pomeron is correctly
treated as a $C= +1$ exchange, as is the case for
our tensor pomeron. On the other hand, in a vector-pomeron model,
using standard QFT rules for the vertices, we will have effectively
a $C = -1$ pomeron.
Then, all exchanges will be, effectively,
$(C_{1},C_{2}) = (-1,-1)$ and all asymmetries
(\ref{asymmetry_eta}) to (\ref{asymmetry_3}) will be zero.
We cannot and do not exclude the possibility that by
introducing some \textit{ad hoc} sign changes in amplitudes one can generate
non-zero asymmetries also in vector-pomeron models.
But we emphasize that in the tensor-pomeron model \cite{Ewerz:2013kda}
asymmetries are generated in a natural and straightforward way.
Thus, experimental observations of such asymmetries would give
strong support for the tensor-pomeron concept.

\begin{figure}[!ht]
\includegraphics[width=0.48\textwidth]{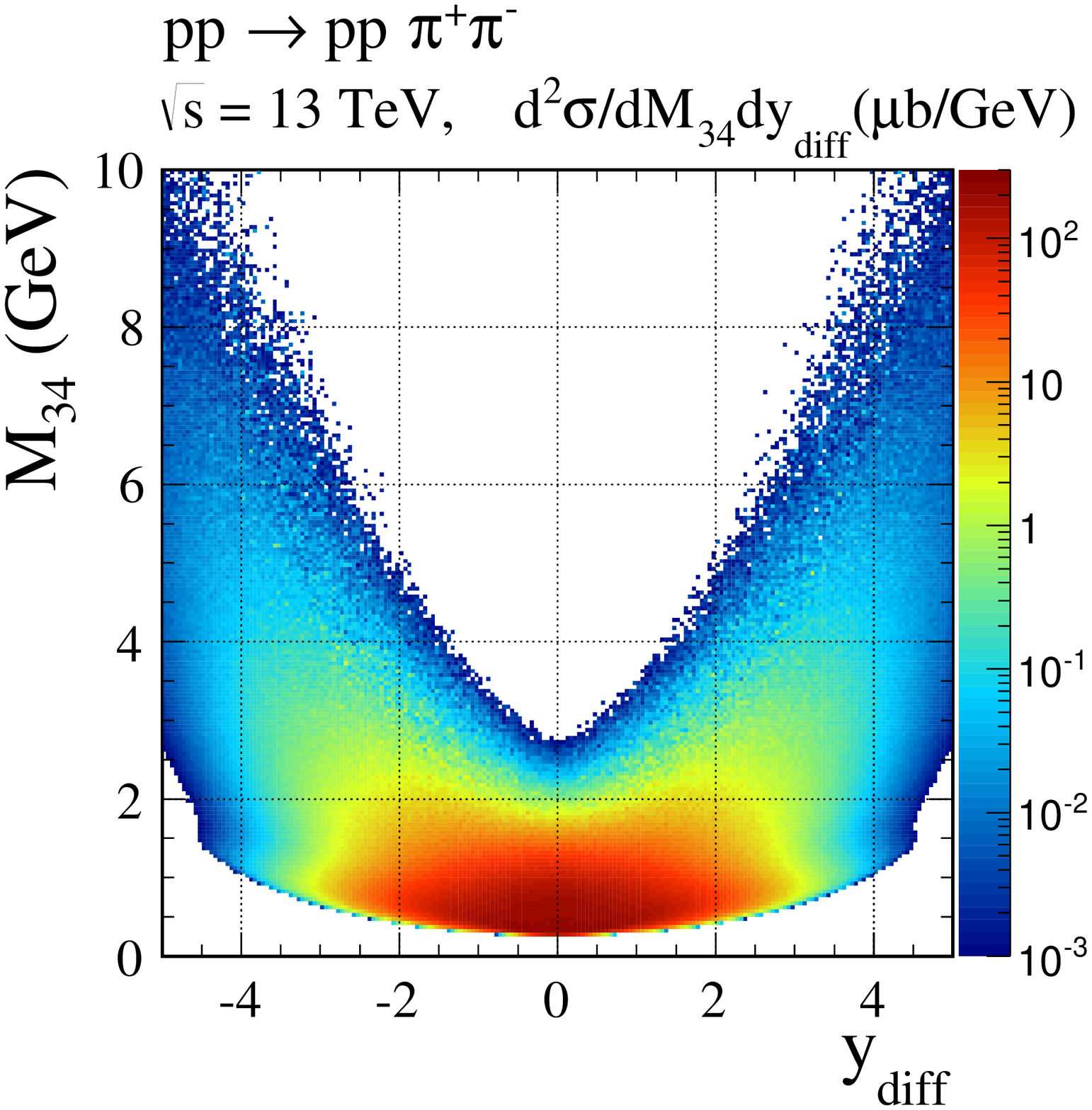}
\includegraphics[width=0.48\textwidth]{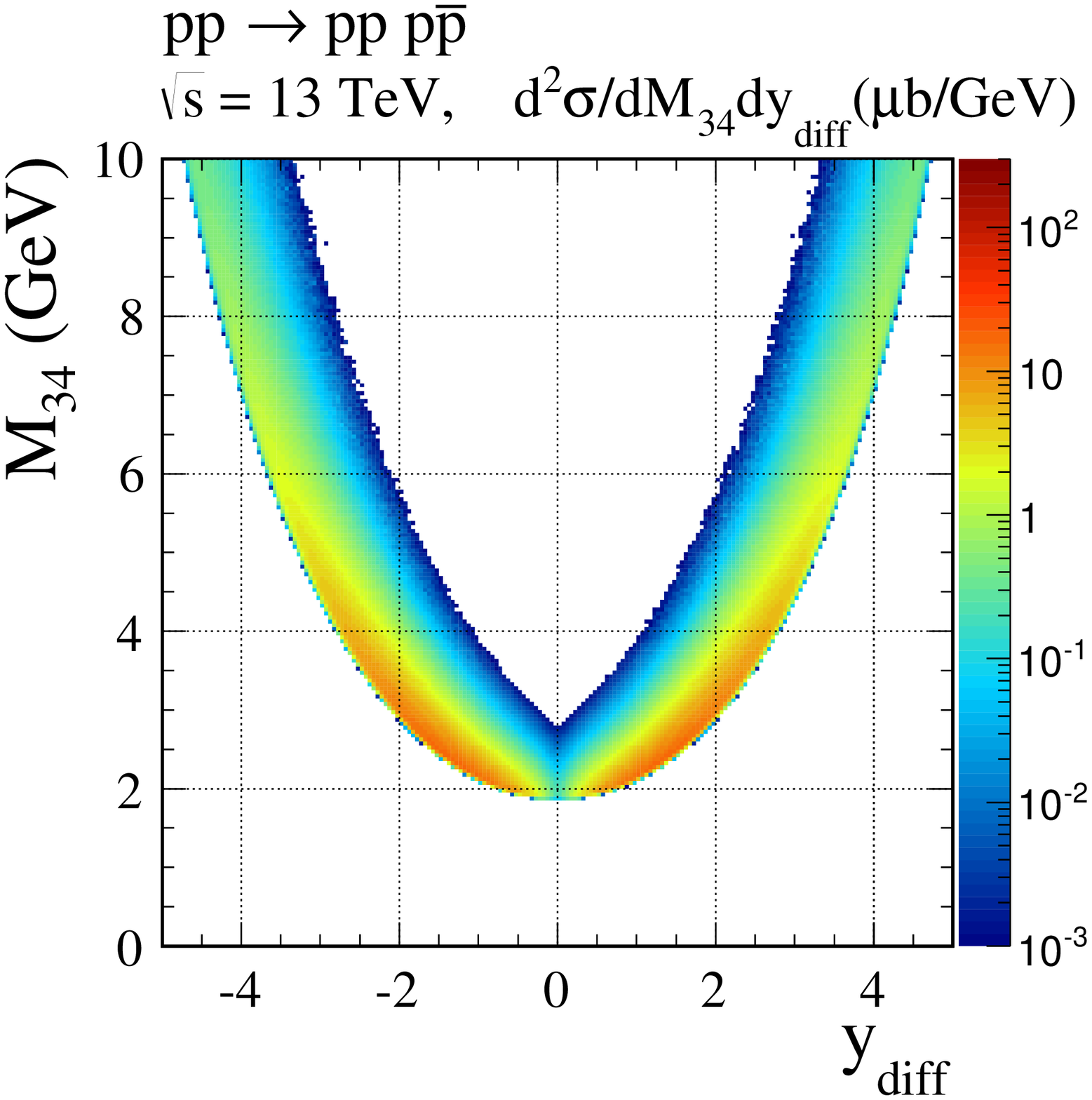}
  \caption{\label{fig:dsig_dM34dydiff}
  \small
The two-dimensional distributions in ($M_{34}, \rm{y}_{diff}$) 
for the diffractive continuum $\pi^{+}\pi^{-}$ (the left panel) 
and $p\bar{p}$ (the right panel) production for the full phase space at $\sqrt{s} = 13$~TeV.
Results for the combined tensor-pomeron and reggeon exchanges are shown.
We have taken here $\Lambda_{off,E} = 1$~GeV.
No absorption effects have been included here.}
\end{figure}

\begin{figure}[!ht]
\includegraphics[width=0.48\textwidth]{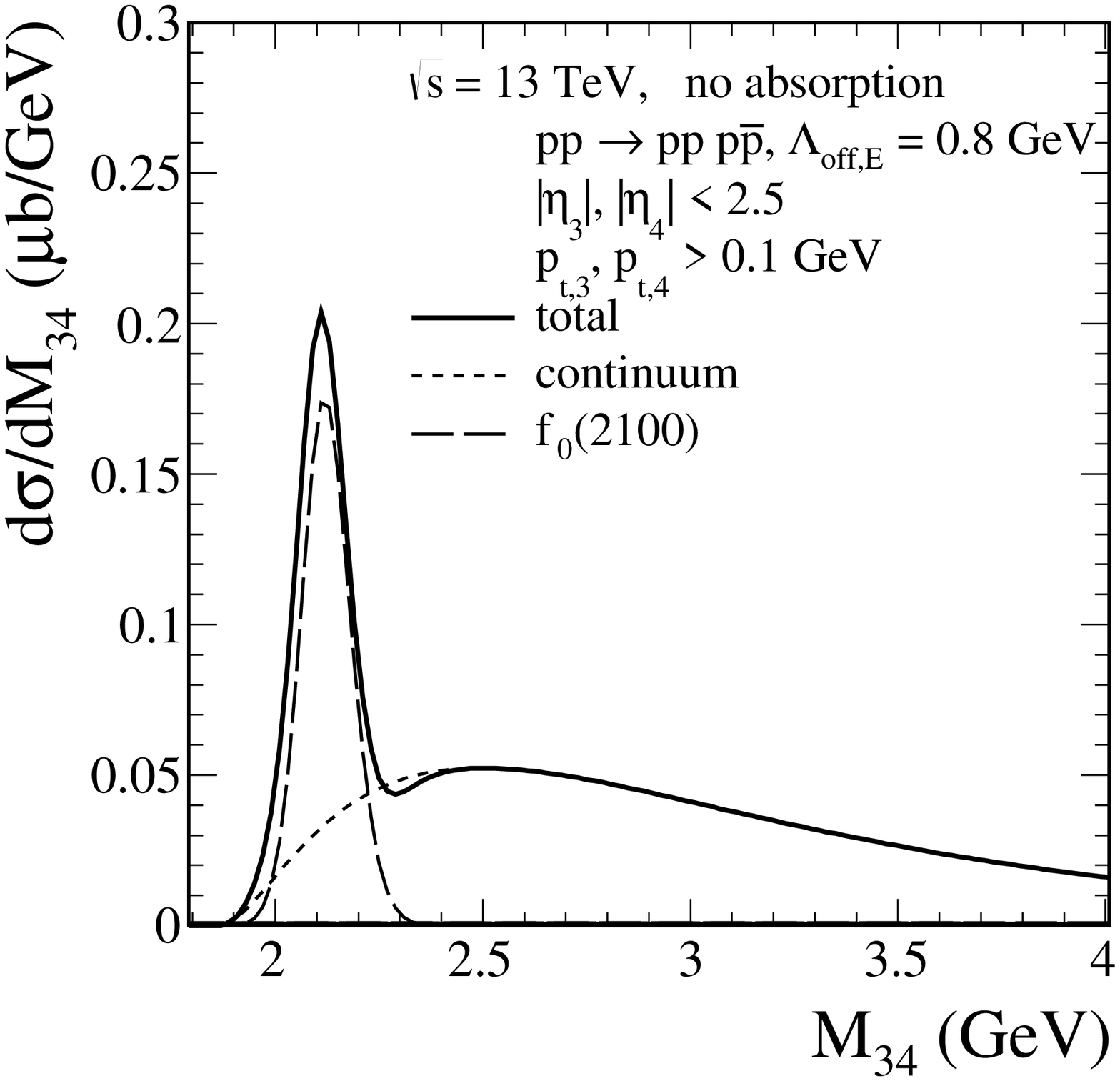}
\includegraphics[width=0.48\textwidth]{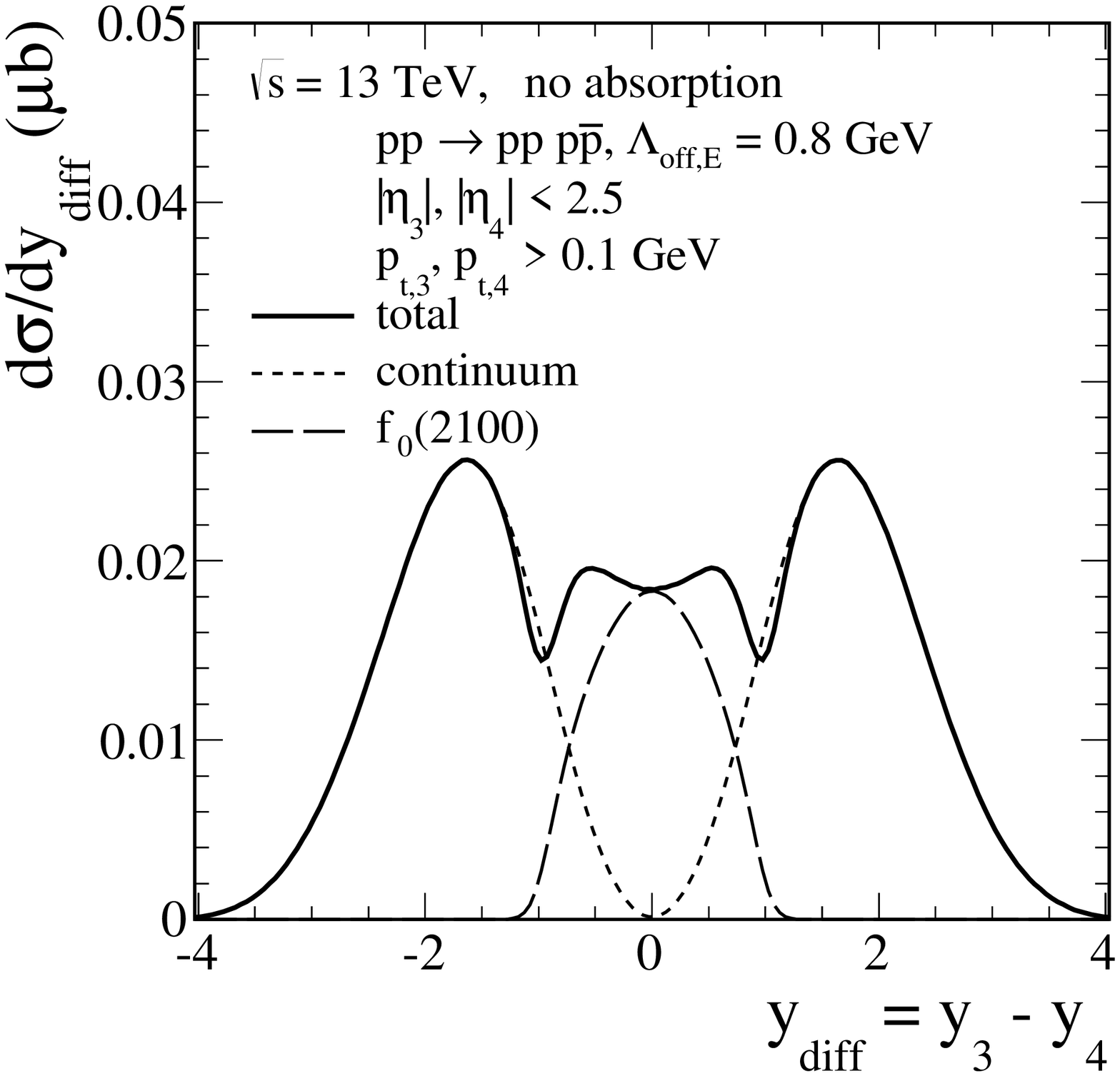}
  \caption{\label{fig:f0}
  \small
The differential cross sections for $pp \to pp p \bar{p}$ at $\sqrt{s} = 13$~TeV
for continuum plus $f_{0}(2100)$ production.
The distributions in the $p \bar{p}$ invariant mass
(the left panel) and in ${\rm y}_{diff}$ (the right panel) are shown.
No absorption effects were included here.}
\end{figure}

Now, we turn to $p \bar{p}$ production via resonances.
Not much is known about mesonic resonances in the $p \bar{p}$ channel,
especially for those produced in the diffractive processes.
Exceptions may be production of $\eta_c$ and $\chi_c$ mesons for which 
the branching fractions to the $p \bar{p}$ channel 
are relatively well known \cite{Olive:2016xmw}.
There is also some evidence for the presence of the $f_{2}(1950)$ resonance
in the $\gamma \gamma \to p \bar{p}$ reaction \cite{Klusek-Gawenda:2017lgt}.
Although statistics of the ISR data \cite{Akesson:1985rn,Breakstone:1989ty} 
was poor for the $pp \to pp p\bar{p}$ reaction,
the data show a large low-mass enhancement.
With good statistics one could study at the LHC the distribution
$d^{2}\sigma/dM_{34} d{\rm y}_{diff}$ 
for the $pp \to pp p \bar{p}$ reaction.
In the right panel of Fig.~\ref{fig:dsig_dM34dydiff},
we show this distribution for the non-resonant $p \bar{p}$ production.
For comparison, in the left panel of Fig.~\ref{fig:dsig_dM34dydiff},
the distribution for the $pp \to pp \pi^{+} \pi^{-}$ reaction is shown.
For $p \bar{p}$ production one can observe a characteristic ridge 
at the edge of the $(M_{34}, {\rm y}_{diff})$ space.
The interior is then free of the diffractive continuum.
There, the identification of possible resonances should be easier.
In reality, the presence of resonances may destroy the dip as resonances 
are expected to give a dominant contribution just at ${\rm y}_{diff} = 0$.

In Fig.~\ref{fig:f0}, we discuss one possible scenario
for the $pp \to pp p \bar{p}$ reaction.
We take into account the non-resonant continuum
including both pomeron and reggeon exchanges
and, as an example, the scalar $f_{0}(2100)$ resonance created by the pomeron-pomeron fusion.
The scalar $f_0(2100)$ was observed in $p \bar{p}$ annihilation into the $\eta \eta$ channel 
using a partial wave analysis of Crystal Barrel data
\cite{Anisovich:2011bw,Anisovich:2000af}.
It may be considered as a second scalar glueball, probably mixed with $q \bar{q}$ states.
For the continuum term, we take $\Lambda_{off,E} = 0.8$~GeV in~(\ref{ff_exp}),
while for the resonant term we take $\Lambda_{f_{0}} = 1$~GeV in~(\ref{Fpompommeson_ff})
and $g_{\Pom \Pom f_{0}}' \, g_{f_{0} p \bar{p}} = 0.8$, $g_{\Pom \Pom f_{0}}''= 0$;
see (\ref{vertex_pompomS}) and Appendix~A of \cite{Lebiedowicz:2013ika}.
Here, the coupling constants are fixed arbitrarily.
We only want to give an example for the effects to be expected
from resonance contributions.
We show the distributions in the $p \bar{p}$ invariant mass
(the left panel) and in ${\rm y}_{diff}$ (the right panel) at $\sqrt{s} = 13$~TeV.
Clearly, the resonant contribution
leads to enhancements at low $M_{p \bar{p}}$ 
and in the central region of ${\rm y}_{diff}$.
We can see that the complete result indicates an interference effect 
of the continuum and $f_{0}(2100)$ terms.
With the parameters used here we get for the complete cross section 113~nb 
for the ATLAS cuts
($|\eta_{3}|,|\eta_{4}| < 2.5$, $p_{t,3}, p_{t,4}> 0.1$~GeV)
and 35~nb for the LHCb cuts 
($2 < \eta_{3}, \eta_{4} < 4.5$, $p_{t,3}, p_{t,4} > 0.2$~GeV)
on centrally produced $p \bar{p}$.
Here the absorption effects are not included.
It is worth adding that the cross section 
for the resonant contribution is concentrated along the diagonal
${\rm y}_{3} \simeq {\rm y}_{4}$ in (${\rm y}_{3}, {\rm y}_{4}$) space,
exactly in the valley of the continuum contribution
(see the right panel in Fig.~\ref{fig:dsig_dy3dy4}). 

\begin{figure}[!ht]
\includegraphics[width=0.5\textwidth]{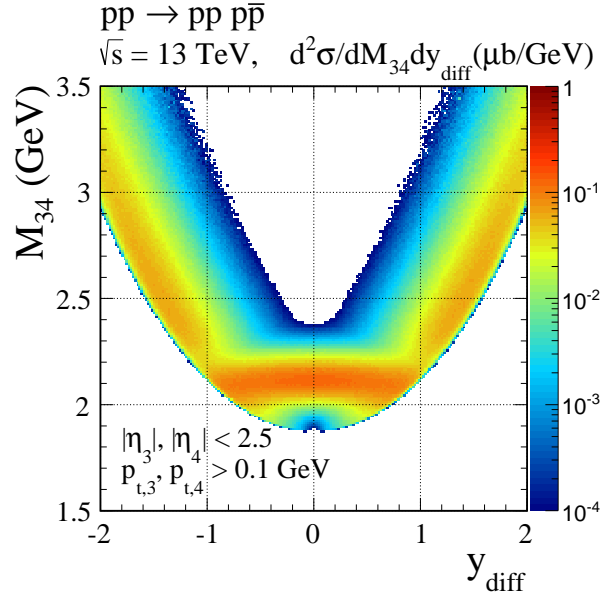}
  \caption{\label{fig:map_M34ydiff_ppbar_f0}
  \small
The two-dimensional distribution in ($M_{34}, \rm{y}_{diff}$) 
for the $p\bar{p}$ production at $\sqrt{s} = 13$~TeV
including the continuum and the $f_{0}(2100)$ contribution.
The strength of the resonance contribution is chosen arbitrarily.
No absorption effects were included here.}
\end{figure}

In Fig.~\ref{fig:map_M34ydiff_ppbar_f0} we show two-dimensional distribution
in $(M_{34}, {\rm y}_{diff})$ for $pp \to pp p \bar{p}$ obtained
from the non-resonant plus the $f_{0}(2100)$ resonant contributions.
Here, the model parameters were chosen as in Fig.~\ref{fig:f0}.
Comparing with the right panel of Fig.~\ref{fig:dsig_dM34dydiff}
we see clearly that the resonance contribution is centered around
$M_{34} = 2.1$~GeV and is approximately uniform in ${\rm y}_{diff}$ 
for $|{\rm y}_{diff}| \lesssim 1$.
Note that for $M_{34} \to 2 m_{p}$, that is,
for $\vec{p}_{3} - \vec{p}_{4} \to 0$ both,
the dominant $(\Pom, \Pom)$ continuum contribution,
as well as the $f_{0}(2100)$ resonance contribution must vanish; 
see (\ref{Tamplitude_aux}).
This is clearly seen in Fig.~\ref{fig:map_M34ydiff_ppbar_f0}.

Also, azimuthal correlations are interesting for central exclusive $p \bar{p}$ production.
From the experimental point of view, this typically would require that the momenta
of the leading protons are measured.
Then, one could study, for instance, the distributions in the angle $\phi_{12}$
between the transverse momenta $\vec{p}_{t,1}$ and $\vec{p}_{t,2}$ 
of the leading protons.
For low-energy central-meson production these angular distributions
have been extensively discussed in \cite{Barberis:1996iq,Barberis:1997ve,Barberis:1998ax,
Barberis:1998sr,Barberis:1999cq,Barberis:2000em,Kirk:2000ws},
\cite{Close:1999is,Close:1999bi,Close:2000dx},
and from the tensor pomeron point of view in \cite{Lebiedowicz:2013ika}.
Angular distributions for glueball production have been discussed in \cite{Iatrakis:2016rvj}.
Since we have constructed in the present paper a model
for central $p \bar{p}$ production at the amplitude level
we could also discuss such azimuthal correlations.
We have checked that our model, including both the continuum and the scalar resonance $f_{0}(2100)$,
taking into account only the $(l,S) = (0,0)$ coupling in (\ref{vertex_pompomS})
gives a rather flat $\phi_{12}$ distribution, 
unlike for central $\pi^{+} \pi^{-}$ production \cite{Lebiedowicz:2014bea,Lebiedowicz:2016ioh}.
This is consistent with a measurement made by the WA102 Collaboration;
see Fig.~4~(a) of \cite{Barberis:1998sr}.
But since present LHC experiments are not yet equipped for such measurements,
we leave this for a further publication.

\section{Conclusions}
\label{sec:conclusions}

In the present article, we have discussed
exclusive production of $p \bar{p}$ and $\Lambda \overline{\Lambda}$
pairs in proton-proton collisions. 
At the present stage, we have taken into account 
mainly the diffractive production of the $p \bar{p}$ continuum.
The amplitudes have been calculated using Feynman rules within
the tensor-pomeron model \cite{Ewerz:2013kda} 
and taking into account the spins of the produced particles.
Applying this model to our reactions here we had to introduce
some form factors containing suitable cut-off parameters;
see (\ref{ff_exp}) and (\ref{Fpompommeson_ff}).
A first estimate of these cut-off parameters was made by
comparing to low-statistics ISR data in which mostly the integrated cross section
for $pp \to pp p \bar{p}$ was measured at $\sqrt{s} = 62$~GeV.
There, we need a cut-off parameter $\Lambda_{off,E} \sim 1$~GeV
in (\ref{ff_exp}).
The form factors and corresponding cut-off parameters needed to describe
the off-shellness of the intermediate $t$-/$u$-channel protons
are not well known and have to be fitted in the future to experimental data. 
They influence mostly the absolute normalization of the cross sections
and have almost no influence on shapes of distributions.
In this paper we did not concentrate on the absolute normalization
but rather on relative effects by
studying the qualitative features of the $pp \to pp p \bar{p}$ reaction
in the tensor-pomeron model.
To describe the relatively low-energy
ISR and WA102 experiments, we find that we have
to include also subleading reggeon exchanges 
in addition to the two-pomeron exchange. 

For our predictions for the LHC we have used 
the off-shell proton form factor parameter in (\ref{ff_exp}),
$\Lambda_{off,E}$, in the range between 0.8 and~1 GeV.
The invariant mass distribution for $p \bar{p}$ pairs is predicted to
extend to larger dihadron invariant masses
than for the production of $\pi^+ \pi^-$ or $K^+ K^-$ or
artificial pseudoscalar nucleons.
This is strongly related to spin 1/2 
for nucleons versus spin 0 for pseudoscalar mesons.

Especially interesting is the distribution in the rapidity difference
between antiproton and proton.
For continuum $p \bar{p}$ production, we predict a dip at ${\rm y}_{diff} = 0$, 
in contrast to $\pi^+ \pi^-$ and $K^+ K^-$ production 
in which a maximum of the cross section occurs at ${\rm y}_{diff} = 0$.
The dip is caused by a good separation of ${\rm \hat{t}}$ and ${\rm \hat{u}}$ contributions 
in (${\rm y}_{3}, {\rm y}_{4}$) space and destructive interference of them
along the diagonal $\rm{y}_{3} = \rm{y}_{4}$ 
characteristic for our Feynman diagrammatic 
calculation with correct treatment of spins.

In our calculations, we have included both pomeron and reggeon exchanges.
The reggeon exchange contributions lead to enhancements
at large absolute values of the $p$ and $\bar{p}$ (pseudo)rapidities.
A similar effect was predicted 
for the $pp \to pp \pi^{+} \pi^{-}$ reaction in \cite{Lebiedowicz:2010yb}.
We have predicted asymmetries in the (pseudo)rapidity distributions of
the centrally produced antiproton and proton.
The asymmetry is caused by interference effects 
of the dominant $(\Pom, \Pom)$ with the subdominant 
$(\Reg_{-}, \Pom + \Reg_{+})$ and $(\Pom + \Reg_{+}, \Reg_{-})$ exchanges.
It should be emphasized that limited detector
acceptances in experimental searches at the LHC 
might affect the size of the asymmetry.
The asymmetry should be much more visible for the LHCb experiment
which covers a region of larger pseudorapidities 
where the reggeon exchanges become more relevant.
Also the odderon will contribute to such asymmetries.
However, we find for typical odderon parameters 
allowed by recent $pp$ elastic data \cite{Antchev:2017yns}
only very small effects,
roughly a factor 10 smaller than the effects due to reggeons 
as predicted in the present paper.

All our predictions here have been done for the tensor-pomeron model.
In the literature, often, a vector pomeron is used, which is
-- strictly speaking -- inconsistent with the rules of quantum field theory
as it gives the pomeron charge conjugation $C = -1$
instead of $C = +1$.
This is discussed, e.g., in Refs.~\cite{Ewerz:2013kda,Lebiedowicz:2016ioh,Ewerz:2016onn}.
Although the vector-pomeron model is incorrect from the field theory
point of view, it leads to almost the same distributions
including the prediction of the dip at ${\rm y}_{diff} = 0$.
This is not too surprising since the leading $(\Pom, \Pom)$ fusion
term has $(C_{1}, C_{2}) = (1,1)$ for the tensor pomeron
and $(C_{1}, C_{2}) = (-1,-1)$ for the vector pomeron, giving in
both cases a state with $C = +1$.
The situation is quite different for pomeron-reggeon,
$(\Pom, \Reg_{-})$ and $(\Reg_{-}, \Pom)$, exchange.
There, we get with a tensor pomeron a $C = -1$ state,
with a vector pomeron again a $C = +1$ state.
The interference of $p \bar{p}$ amplitudes with
$C = +1$ and $C = -1$ leads to the asymmetries 
discussed in Sec.~\ref{sec:results}.
We see great difficulties producing such asymmetries
in a vector-pomeron model in which only $C = +1$ $p \bar{p}$ amplitudes occur.
Therefore, we find it an important task for experimentalist
to study the asymmetries (\ref{asymmetry_eta}) - (\ref{asymmetry_3}).
If non-zero asymmetries are found we would have a further strong
argument in favour of the tensor-pomeron concept.

In the present study, we have focused mainly on the
production of continuum $p \bar{p}$ pairs
in the framework of the tensor-pomeron model
treating correctly the spin degrees of freedom.
Not much is known about diffractively produced $p \bar{p}$ resonances.
Any experimentally observed distortions from 
our continuum-$p \bar{p}$ predictions 
may therefore signal the presence of resonances. 
This could give new interesting information for meson spectroscopy.
We have discussed a first qualitative attempt to ``reproduce''
the experimentally observed behaviors of the $p \bar{p}$ invariant mass ($M_{34}$) spectra
observed in \cite{Akesson:1985rn,Breakstone:1989ty,Barberis:1998sr}.
Our calculation shows that the diffractive production of $p \bar{p}$ 
through the $s$-channel $f_{0}(2100)$ resonance
leads to an enhancement at low $M_{34}$ and
that the resonance contribution is concentrated at $|{\rm y}_{diff}| < 1$.
In general, more resonances can contribute, 
e.g., $f_{0}(2020)$, $f_{0}(2200)$, and $f_{0}(2300)$.
Contributions of other states, 
such as $f_{2}(1950)$, are not excluded.
Also, the subthreshold $m_R < 2 m_p$ resonances that 
would effectively generate a continuum $p \bar{p}$ contribution
should be taken into account; see \cite{Klusek-Gawenda:2017lgt}.
Interference effects between the continuum and resonant mechanisms 
certainly will occur; see Fig.~\ref{fig:f0}.

The predictions made for $p \bar{p}$ production can be easily repeated 
for diffractive $\Lambda \overline{\Lambda}$ pair production. 
Here, the uncertainties for the continuum contribution 
are slightly larger than for the $p \bar{p}$ production 
(higher off-shell effects and less-known interaction parameters).
However, here, the resonance contributions 
are expected to be much smaller if present at all. 
Any clear observation of a resonance 
in the $\Lambda \overline{\Lambda}$ channel would, therefore, 
be a sensation, and the result would definitely go to the Particle Data Book.
On the other hand, a lack of such resonances would allow a verification
of the minimum at ${\rm y}_{diff} = 0$, which we predict 
using the correct treatment of the spin degrees of freedom 
in the Regge-like calculations of central exclusive baryon-antibaryon production.

\acknowledgments
The authors are grateful to Leszek Adamczyk, Carlo Ewerz,
and Rainer Schicker for discussions.
This work was partially supported by
the Polish National Science Centre Grant No. 2014/15/B/ST2/02528
and by the Center for Innovation and Transfer of Natural Sciences 
and Engineering Knowledge in Rzesz\'ow.

\appendix

\section{Effective propagators and vertices for pomeron, reggeon, and odderon exchange}
\label{sec:appendixA}

Here, we collect the expressions for our effective exchanges and vertex functions as given
in Sec.~3 of \cite{Ewerz:2013kda} in order to make our present paper self-contained.
For extensive discussions motivating the following expressions, we refer to \cite{Ewerz:2013kda}.

Our effective pomeron propagator reads
\begin{eqnarray}
i \Delta^{(\Pom)}_{\mu \nu, \kappa \lambda}(s,t) = 
\frac{1}{4s} \left( g_{\mu \kappa} g_{\nu \lambda} 
                  + g_{\mu \lambda} g_{\nu \kappa}
                  - \frac{1}{2} g_{\mu \nu} g_{\kappa \lambda} \right)
(-i s \alpha'_{\Pom})^{\alpha_{\Pom}(t)-1}
\label{A1}
\end{eqnarray}
and fulfills the following relations:
\begin{equation} 
\begin{split}
&\Delta^{(\Pom)}_{\mu \nu, \kappa \lambda}(s,t) = 
\Delta^{(\Pom)}_{\nu \mu, \kappa \lambda}(s,t) =
\Delta^{(\Pom)}_{\mu \nu, \lambda \kappa}(s,t) =
\Delta^{(\Pom)}_{\kappa \lambda, \mu \nu}(s,t) \,, \\
&g^{\mu \nu} \Delta^{(\Pom)}_{\mu \nu, \kappa \lambda}(s,t) = 0, \quad 
g^{\kappa \lambda} \Delta^{(\Pom)}_{\mu \nu, \kappa \lambda}(s,t) = 0 \,.
\end{split}
\label{A2}
\end{equation}
%
Here, the pomeron trajectory $\alpha_{\Pom}(t)$
is assumed to be of standard linear form, see e.g. \cite{Donnachie:2002en},
\begin{eqnarray}
&&\alpha_{\Pom}(t) = \alpha_{\Pom}(0)+\alpha'_{\Pom}\,t\,, \nonumber\\
&&\alpha_{\Pom}(0) = 1.0808\,, \nonumber\\
&&\alpha'_{\Pom} = 0.25 \; \mathrm{GeV}^{-2}\,.
\label{A3}
\end{eqnarray}
The pomeron-proton vertex function, supplemented by a vertex form factor, 
taken here to be the Dirac electromagnetic form factor of the proton for simplicity,
has the form
%
\begin{eqnarray}
&&i\Gamma_{\mu \nu}^{(\Pom pp)}(p',p)= 
i\Gamma_{\mu \nu}^{(\Pom \bar{p} \bar{p})}(p',p)
\nonumber\\
&& \quad =-i 3 \beta_{\Pom NN} F_{1}\bigl((p'-p)^2\bigr)
\left\lbrace 
\frac{1}{2} 
\left[ \gamma_{\mu}(p'+p)_{\nu} 
     + \gamma_{\nu}(p'+p)_{\mu} \right]
- \frac{1}{4} g_{\mu \nu} ( p\!\!\!/' + p\!\!\!/ )
\right\rbrace\,, \qquad
\label{A4}
\end{eqnarray}
with 
$\beta_{\Pom NN} = 1.87$~GeV$^{-1}$.

The ansatz for the $C=+1$ reggeons $\Reg_{+} = f_{2 \Reg}, a_{2 \Reg}$
is similar to (\ref{A1}) - (\ref{A4}).
The $\Reg_{+}$ propagator is obtained from (\ref{A1}) with the replacements
\begin{eqnarray}
&&\alpha_{\Pom}(t) \to \alpha_{\Reg_{+}}(t) = \alpha_{\Reg_{+}}(0)+\alpha'_{\Reg_{+}}\,t\,, \nonumber \\
&&\alpha_{\Reg_{+}}(0) = 0.5475\,, \nonumber \\
&&\alpha'_{\Reg_{+}} = 0.9 \; \mathrm{GeV}^{-2}\,.
\label{A5}
\end{eqnarray}
The $f_{2 \Reg}$- and $a_{2 \Reg}$-proton vertex functions are obtained from (\ref{A4})
with the replacements ($M_{0} = 1$~GeV)
\begin{eqnarray}
&&3 \beta_{\Pom NN} \to \frac{g_{f_{2 \Reg} pp}}{M_{0}}\,, \nonumber \\
&&g_{f_{2 \Reg} pp} = 11.04\,,
\label{A6}
\end{eqnarray}
and
\begin{eqnarray}
&&3 \beta_{\Pom NN} \to \frac{g_{a_{2 \Reg} pp}}{M_{0}}\,, \nonumber \\
&&g_{a_{2 \Reg} pp} = 1.68\,,
\label{A7}
\end{eqnarray}
respectively.

Our ansatz for the $C=-1$ reggeons $\Reg_{-} = \omega_{\Reg}, \rho_{\Reg}$
reads as follows. We assume an effective vector propagator
\begin{eqnarray}
i \Delta^{(\Reg_{-})}_{\mu \nu}(s,t) = 
i g_{\mu \nu} \frac{1}{M_{-}^{2}} (-i s \alpha'_{\Reg_{-}})^{\alpha_{\Reg_{-}}(t)-1}\,,
\label{A8}
\end{eqnarray}
with
\begin{eqnarray}
&&\alpha_{\Reg_{-}}(t) = \alpha_{\Reg_{-}}(0)+\alpha'_{\Reg_{-}}\,t\,, \nonumber \\
&&\alpha_{\Reg_{-}}(0) = 0.5475\,, \nonumber \\
&&\alpha'_{\Reg_{-}} = 0.9 \; \mathrm{GeV}^{-2}\,,\nonumber \\
&&M_{-} = 1.41 \; \mathrm{GeV}\,.
\label{A9}
\end{eqnarray}
The $\Reg_{-}$-proton vertex reads ($\Reg_{-} = \omega_{\Reg}, \rho_{\Reg}$)
\begin{eqnarray}
i\Gamma_{\mu}^{(\Reg_{-} pp)}(p',p)= -i\Gamma_{\mu}^{(\Reg_{-} \bar{p} \bar{p})}(p',p)
                                   = -i g_{\Reg_{-} pp} F_{1}\bigl((p'-p)^2\bigr) \gamma_{\mu}\,,
\label{A10}
\end{eqnarray}
with
\begin{eqnarray}
&&g_{\omega_{\Reg} pp} = 8.65\,, \nonumber \\
&&g_{\rho_{\Reg} pp} = 2.02\,.
\label{A11}
\end{eqnarray}

Our ansatz for the odderon follows (3.16), (3.17) and (3.68), (3.69) of \cite{Ewerz:2013kda}:
\begin{eqnarray}
&&i \Delta^{(\Ode)}_{\mu \nu}(s,t) = 
-i g_{\mu \nu} \frac{\eta_{\Ode}}{M_{0}^{2}} (-i s \alpha'_{\Ode})^{\alpha_{\Ode}(t)-1}\,,
\label{A12} \\
&&i\Gamma_{\mu}^{(\Ode pp)}(p',p) = -i\Gamma_{\mu}^{(\Ode \bar{p} \bar{p})}(p',p)
                                  = -i 3\beta_{\Ode pp} M_{0}\,F_{1}\bigl((p'-p)^2\bigr) \gamma_{\mu}\,.
\label{A13}
\end{eqnarray}
We take here what we think are representative values
for the odderon parameters in light of the recent TOTEM results \cite{Antchev:2017yns},
\begin{eqnarray}
&&\eta_{\Ode} = -1\,, \nonumber \\
&&\alpha_{\Ode}(t) = \alpha_{\Ode}(0)+\alpha'_{\Ode}\,t\,, \nonumber \\
&&\alpha_{\Ode}(0) = 1.05\,, \nonumber \\
&&\alpha'_{\Ode} = 0.25 \; \mathrm{GeV}^{-2}\,, \nonumber\\
&&\beta_{\Ode NN} = 0.2 \; \mathrm{GeV}^{-1}\,.
\label{A14}
\end{eqnarray}
All numbers for the parameters listed above should be considered as default values to be checked
and -- if necessary -- adjusted using relevant experimental data.
Some estimates of the present uncertainties of the parameters are discussed in Sec.~3 of Ref.~\cite{Ewerz:2013kda}.

\bibliography{refs}

\end{document}